\documentclass[oldversion]{aa}
\usepackage{graphicx}
\usepackage{txfonts}
\usepackage{natbib}
\bibpunct{(}{)}{;}{a}{}{,}
\begin{document}

\title{The transverse proximity effect in spectral hardness on the line of sight towards HE~2347$-$4342
  \thanks{Based on observations collected
    at the European Southern Observatory, Chile (Proposals 070.A-0425 and
    074.A-0273). Data collected under Proposals 068.A-0194, 070.A-0376 and 116.A-0106 was obtained from
    the ESO Science Archive.}
} 

\titlerunning{The transverse proximity effect in spectral hardness towards HE~2347$-$4342}
\author{G.~Worseck \inst{1} \and C.~Fechner \inst{2,3} \and L.~Wisotzki \inst{1} \and A.~Dall'Aglio \inst{1}}
\institute{Astrophysikalisches Institut Potsdam, An der Sternwarte 16, 14482 Potsdam, Germany \and
Hamburger Sternwarte, Universit\"{a}t Hamburg, Gojenbergsweg 112, 21029 Hamburg, Germany \and
Universit\"{a}t Potsdam, Am Neuen Palais 10, 14469 Potsdam, Germany}
\offprints{G.~Worseck, \email{gworseck@aip.de}}
\date{Received 2 April 2007 / Accepted 31 July 2007}

\abstract{We report the discovery of 14 quasars in the vicinity of
HE~2347$-$4342, one of the two quasars whose intergalactic \ion{He}{ii}
forest has been resolved with FUSE. By analysing the \ion{H}{i} and the
\ion{He}{ii} opacity variations separately, no transverse proximity effect is
detected near three foreground quasars of HE~2347$-$4342:
QSO~J23503$-$4328 ($z=2.282$, $\vartheta=3\farcm59$), QSO~J23500$-$4319
($z=2.302$, $\vartheta=8\farcm77$) and QSO~J23495$-$4338
($z=2.690$, $\vartheta=16\farcm28$). This is primarily due to line
contamination and overdensities probably created by large-scale structure.
By comparing the \ion{H}{i} absorption and the corresponding \ion{He}{ii} absorption, we
estimated the fluctuating spectral shape of the extragalactic UV radiation field
along this line of sight. We find that the UV spectral shape near HE~2347$-$4342
and in the projected vicinity of the three foreground quasars is statistically
harder than expected from UV background models dominated by quasars.
In addition, we find three highly ionised metal line systems near the quasars.
However, they do not yield further constraints on the shape of the ionising field.
We conclude that the foreground quasars show a transverse proximity effect that
is detectable as a local hardening of the UV radiation field, although the
evidence is strongest for QSO~J23495$-$4338. Thus, the relative spectral
hardness traces the proximity effect also in overdense regions prohibiting the
traditional detection in the \ion{H}{i} forest. Furthermore, we emphasise that
softening of quasar radiation by radiative transfer in the intergalactic medium
is important to understand the observed spectral shape variations. From the
transverse proximity effect of QSO~J23495$-$4338 we obtain a lower limit on the
quasar lifetime of $\sim 25$~Myr.

} 

\maketitle

\section{Introduction}

After reionisation the intergalactic medium (IGM) is kept highly photoionised by
the metagalactic UV radiation field generated by the overall population of
quasars and star-forming galaxies
\citep[e.g.][]{haardt96,fardal98,bianchi01,sokasian03}. The intensity and
spectral shape of the UV background determines the ionisation state of the
observable elements in the IGM. In particular, the remaining 
fraction of intergalactic neutral hydrogen and singly ionised helium is
responsible for the Ly$\alpha$ forest of \ion{H}{i} and \ion{He}{ii}.

On lines of sight passing near quasars the IGM will be statistically more
ionised due to the local enhancement of the UV flux that should result in a
statistically higher IGM transmission ('void') in the QSO's vicinity
\citep{fardal93,croft04,mcdonald05}. This so-called proximity effect has been
found with high statistical significance on lines of sight towards luminous
quasars \citep[e.g.][]{bajtlik88,giallongo96,scott00}. On the other hand, a
transverse proximity effect created by foreground ionising sources nearby the
line of sight has not been clearly detected in the \ion{H}{i} forest, except the
recent detection at $z=5.70$ by \citet{gallerani07}. While two large
\ion{H}{i} voids have been claimed to be due to the transverse proximity effect
by \citet[][however see \citealt{dobrzycki91b}]{dobrzycki91} and
\citet{srianand97}, other studies find at best marginal evidence
\citep{fernandez-soto95,liske01}, and most attempts resulted in non-detections
\citep{crotts89,moller92,crotts98,schirber04,croft04}. This has led to
explanations involving the systematic effects of anisotropic radiation, quasar
variability \citep{schirber04}, intrinsic overdensities
\citep{loeb95,rollinde05,hennawi07,guimaraes07} and finite quasar lifetimes
\citep{croft04}.

Intergalactic \ion{He}{ii} Ly$\alpha$ absorption
($\lambda_\mathrm{rest}=303.7822$~\AA) can be studied only towards the few
quasars at $z>2$ whose far UV flux is not extinguished by intervening Lyman limit
systems \citep{picard93,jakobsen98}. Of the six quasars successfully observed so
far, the lines of sight towards \object{HE~2347$-$4342} ($z=2.885$) and
\object{HS~1700$+$6416} ($z=2.736$) probe the post-reionisation era of
\ion{He}{ii} with an emerging \ion{He}{ii} forest that has been resolved with
FUSE \citep{kriss01,shull04,zheng04,fechner06,fechner07}.

In a highly ionised IGM a comparison of the \ion{H}{i} with the corresponding
\ion{He}{ii} absorption yields an estimate of the spectral shape of the UV
radiation field due to the different ionisation thresholds of both species. The
amount of \ion{He}{ii} compared to \ion{H}{i} gives a measure of the spectral
softness, generally expressed via the column density ratio
$\eta=N_\ion{He}{ii}/N_\ion{H}{i}$. Typically, $\eta\la 100$ indicates a hard
radiation field generated by the surrounding quasar population, whereas
$\eta\ga 100$ requires a significant contribution of star-forming galaxies or
heavily softened quasar radiation \citep[e.g.][]{haardt96,fardal98,haardt01}.

The recent FUSE observations of the \ion{He}{ii} Ly$\alpha$ forest revealed
large $\eta$ fluctuations ($1\la\eta\la 1000$) on small scales of
$0.001\la \Delta z\la 0.03$ with a median $\eta\simeq 80$--100. Apart from
scatter due to the low-quality \ion{He}{ii} data at $S/N\sim 5$
\citep{fechner06,liu06} and possible systematic errors due to the generally
assumed line broadening mechanism \citep{fechner07}, several physical reasons
for these $\eta$ variations have been proposed. A combination of local density
variations \citep{miralda-escude00}, radiative transfer effects
\citep{maselli05,tittley06} and local differences in the properties of quasars
may be responsible for the fluctuations. In particular, at any given point in
the IGM at $z>2$ only a few quasars with a range of spectral indices
\citep{telfer02,scott04} contribute to the UV background at $h\nu\ge 54.4$~eV
\citep{bolton06}.

Already low-resolution \ion{He}{ii} spectra obtained with HST indicate a
fluctuating radiation field, which has been interpreted as the onset of
\ion{He}{ii} reionisation in Str\"{o}mgren spheres around hard \ion{He}{ii}
photoionising sources along or near the line of sight
\citep{reimers97,heap00,smette02}. \citet{jakobsen03} found a quasar coinciding
with the prominent \ion{He}{ii} void at $z=3.05$ towards \object{Q~0302$-$003},
thereby presenting the first clear case of a transverse proximity effect. In
\citet{worseck06}, hereafter Paper~I, we revealed the transverse proximity
effect as a systematic increase in spectral hardness around all four known
foreground quasars along this line of sight. This suggests that a hard radiation
field is a sensitive probe of the transverse proximity effect even if there is
no associated void in the \ion{H}{i} forest, either because of the weakness of
the effect, or because of large-scale structure.

Along the line of sight towards HE~2347$-$4342 several \ion{He}{ii} voids have
been claimed to be due to nearby unknown AGN \citep{smette02}. Likewise, some
forest regions with a detected hard radiation field may correspond
to proximity effect zones of putative foreground quasars \citep{fechner07}. Here
we report on results from a slitless spectroscopic quasar survey in the vicinity
of HE~2347$-$4342 and on spectral shape fluctuations of the UV radiation field
probably caused by foreground quasars towards the sightline of HE~2347$-$4342.
The paper is structured as follows. Sect.~\ref{observations} presents the
observations and the supplementary data employed for the paper. Although we do
not detect any transverse proximity effect in the \ion{H}{i} forest
(Sect.~\ref{trprox_h1}), the fluctuating UV spectral shape along the line of
sight indicates a hard radiation field in the projected vicinity of the
foreground quasars (Sect.~\ref{los_hardness}).
In Sect.~\ref{metalsystem} we study three nearby metal line systems which could
further constrain the ionising field. We interpret the statistically
significant excesses of hard radiation as being due to the transverse proximity
effect (Sect.~\ref{discussion}). We present our conclusions in
Sect.~\ref{conclusions}. Throughout the paper we adopt a flat cosmological model
with $\Omega_\mathrm{m}=0.3$, $\Omega_\Lambda=0.7$ and
$H_0=70$~$\mathrm{km}\,\mathrm{s}^{-1}\,\mathrm{Mpc}^{-1}$.

\section{Observations and data reduction}
\label{observations}

\subsection{Search for QSO candidates near HE~2347$-$4342}

In October 2002 we observed a $25\arcmin\times 33\arcmin$ field centered on
HE~2347$-$4342 ($z=2.885$) with the ESO Wide Field Imager \citep[WFI,][]{baade99}
at the ESO/MPI 2.2~m Telescope (La Silla) in its slitless spectroscopic mode
\citep{wisotzki01} as part of a survey for faint quasars in the vicinity of
established high-redshift quasars. A short summary of the survey is given in
Paper~I; a detailed description will follow in a separate paper.

A semi-automated search for emission line objects among the slitless spectra of
the $\sim 1400$ detected objects in the field resulted in 10 prime quasar
candidates.

\subsection{Spectroscopic follow-up}

\begin{figure*}
\centering
\includegraphics[scale=1.05]{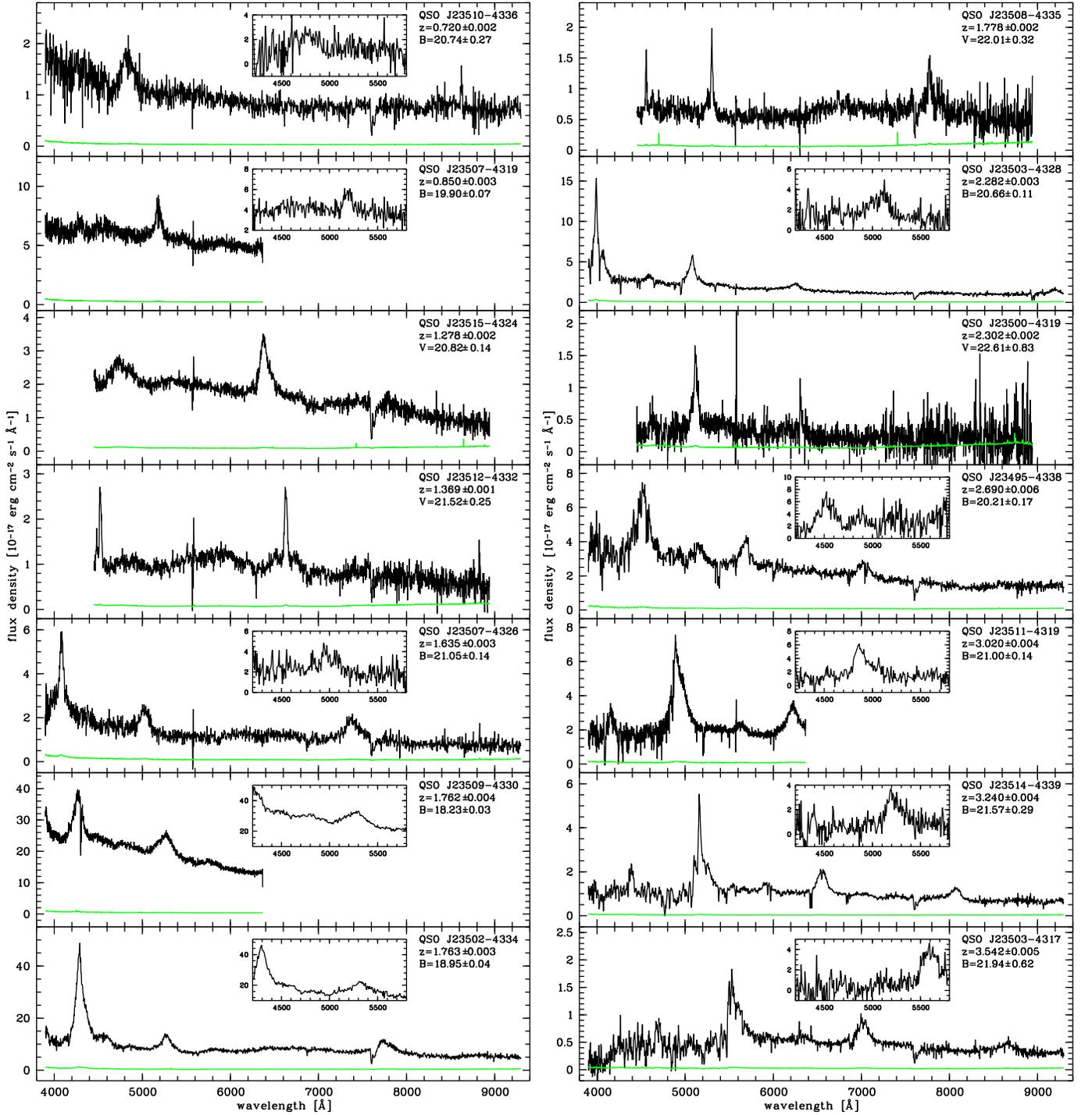}
\caption{\label{he2347qsospec} VLT/FORS spectra of quasars in the vicinity of 
HE~2347$-$4342. The spectra are shown in black together with their $1\sigma$
noise arrays (green lines). The small inserts show the corresponding discovery
spectra from our slitless survey in the same units.}
\end{figure*}

Follow-up spectroscopy of these 10 quasar candidates was obtained with the Focal 
Reducer/Low Dispersion Spectrograph 2 \citep[FORS2,][]{appenzeller98} on
ESO VLT UT1/Antu in Visitor Mode on November 17 and 19, 2004 under variable
seeing but clear conditions. The spectra were taken either with the 300V grism
or the 600B grism and a 1\arcsec\ slit kept at the parallactic angle, resulting
in a spectral resolution of $\sim 10$~\AA\ \textit{FWHM} and $\sim 4.5$~\AA\
\textit{FWHM}, respectively.
No order separation filter was employed, leading to possible order overlap at
$\lambda>6600$~\AA\ in the spectra taken with the 300V grism. Exposure times were
adjusted to yield $S/N\sim 20$ in the quasar continuum. The spectra were
calibrated in wavelength against the FORS2 He/Ne/Ar/HgCd arc lamps and
spectrophotometrically calibrated against the HST standard stars Feige~110 and
GD~108. Data reduction was performed with standard IRAF tasks using the optimal
extraction algorithm by \citet{horne86}. Figure~\ref{he2347qsospec} shows the
spectra of the quasars together with 4 quasars from another survey
(Sect.~\ref{jakobsensurvey}). Table~\ref{observinglog} summarises our
spectroscopic follow-up observations.

\subsection{Additional quasars}
\label{jakobsensurvey}

We checked the ESO Science Archive for additional quasars in the vicinity of
HE~2347$-$4342 and found several unpublished quasars from a deeper slitless 
spectroscopic survey using the ESO VLT, the results of which (on the field of
Q~0302$-$003) are described in \citet{jakobsen03}. We obtained their follow-up
spectra of quasars surrounding HE~2347$-$4342 from the archive and publish them
here in agreement with P.~Jakobsen. In the course of their survey FORS1 spectra
of 10 candidates were taken with the 300V grism crossed with the GG435 order
separation filter and a 1\arcsec\ slit, calibrated against the standards LTT~7987
and GD~50. Seven of their candidates are actually quasars, of which 3 were also
found independently by our survey. The remaining 4 quasars are beyond our
redshift-dependent magnitude limit. The FORS1 spectra of the 4 additional
quasars are displayed in Fig.~\ref{he2347qsospec} and listed separately in
Table~\ref{observinglog}.
According to the quasar catalogue by \citet{veron06} there are no other
previously known quasars within a radius $<30\arcmin$ around HE~2347$-$4342.

\begin{table*}
\caption{Quasars observed near the line of sight of HE~2347$-$4342. The first 10
listed quasars have been found in our survey, the remaining 4 quasars result from
the previously unpublished survey by P.~Jakobsen. Quasar magnitudes are $B$ and
$V$ magnitudes for our survey and Jakobsen's survey, respectively.}
\label{observinglog}
\centering\scriptsize
\begin{tabular}{lllllllllll}
\hline\hline\noalign{\smallskip}
Object &$\alpha$~(J2000) &$\delta$~(J2000) &$z$	&Magnitude&Night &Grism &Exposure &Airmass&Seeing&Abbr.\\
\noalign{\smallskip}\hline\noalign{\smallskip}
\object{QSO~J23510$-$4336}&$23^\mathrm{h}51^\mathrm{m}05\fs50$	&$-43\degr36\arcmin57\farcs2$
&$0.720\pm0.002$	&$20.74\pm0.27$	&19 Nov 2004 	&300V	&1200~s &$1.30$ &1\farcs3	&\\
\object{QSO~J23507$-$4319}&$23^\mathrm{h}50^\mathrm{m}44\fs97$	&$-43\degr19\arcmin26\farcs0$
&$0.850\pm0.003$	&$19.90\pm0.07$	&17 Nov 2004	&600B	&360~s 	&$1.28$ &0\farcs7	&\\
\object{QSO~J23507$-$4326}&$23^\mathrm{h}50^\mathrm{m}45\fs39$	&$-43\degr26\arcmin37\farcs0$
&$1.635\pm0.003$	&$21.05\pm0.14$	&17 Nov 2004	&300V	&200~s  &$1.23$ &1\farcs0	&\\
\object{QSO~J23509$-$4330}&$23^\mathrm{h}50^\mathrm{m}54\fs80$	&$-43\degr30\arcmin42\farcs2$
&$1.762\pm0.004$	&$18.23\pm0.03$	&17 Nov 2004 	&600B	&300~s  &$1.08$ &0\farcs7	&\\
\object{QSO~J23502$-$4334}&$23^\mathrm{h}50^\mathrm{m}16\fs18$	&$-43\degr34\arcmin14\farcs7$
&$1.763\pm0.003$	&$18.95\pm0.04$	&17 Nov 2004 	&300V	&60~s  	&$1.18$ &0\farcs7	&\\
\object{QSO~J23503$-$4328}&$23^\mathrm{h}50^\mathrm{m}21\fs55$	&$-43\degr28\arcmin43\farcs7$
&$2.282\pm0.003$	&$20.66\pm0.11$	&17 Nov 2004	&300V	&400~s 	&$1.20$ &0\farcs7	&A\\
\object{QSO~J23495$-$4338}&$23^\mathrm{h}49^\mathrm{m}34\fs53$	&$-43\degr38\arcmin08\farcs7$
&$2.690\pm0.006$	&$20.21\pm0.17$	&19 Nov 2004	&300V	&360~s  &$1.13$ &1\farcs2	&C\\
\object{QSO~J23511$-$4319}&$23^\mathrm{h}51^\mathrm{m}09\fs44$	&$-43\degr19\arcmin41\farcs6$
&$3.020\pm0.004$	&$21.00\pm0.14$	&17 Nov 2004	&600B	&1000~s &$1.09$ &1\farcs1	&\\
\object{QSO~J23514$-$4339}&$23^\mathrm{h}51^\mathrm{m}25\fs54$	&$-43\degr39\arcmin02\farcs9$
&$3.240\pm0.004$	&$21.57\pm0.29$	&17 Nov 2004	&300V	&1400~s	&$1.14$	&1\farcs2	&\\
\object{QSO~J23503$-$4317}&$23^\mathrm{h}50^\mathrm{m}21\fs94$	&$-43\degr17\arcmin30\farcs0$
&$3.542\pm0.005$	&$21.94\pm0.62$	&19 Nov 2004	&300V	&1800~s &$1.23$ &1\farcs2	&\\
&&&			&	      	&		&600B 	&1800~s &$1.33$ &1\farcs2	&\\\hline
\object{QSO~J23515$-$4324}&$23^\mathrm{h}51^\mathrm{m}33\fs05$	&$-43\degr24\arcmin45\farcs2$
&$1.278\pm0.002$	&$20.82\pm0.14$	&06 Oct 2002 	&300V 	&900~s  &$1.24$ &0\farcs7	&\\
\object{QSO~J23512$-$4332}&$23^\mathrm{h}51^\mathrm{m}15\fs18$	&$-43\degr32\arcmin34\farcs3$
&$1.369\pm0.001$	&$21.52\pm0.25$	&06 Oct 2002	&300V	&900~s  &$1.18$ &0\farcs7	&\\
\object{QSO~J23508$-$4335}&$23^\mathrm{h}50^\mathrm{m}52\fs91$	&$-43\degr35\arcmin06\farcs8$
&$1.778\pm0.002$	&$22.01\pm0.32$	&06 Oct 2002	&300V	&900~s  &$1.11$ &0\farcs9	&\\
\object{QSO~J23500$-$4319}&$23^\mathrm{h}50^\mathrm{m}00\fs28$	&$-43\degr19\arcmin46\farcs1$
&$2.302\pm0.002$	&$22.61\pm0.83$	&06 Oct 2002	&300V 	&900~s 	&$2.37$ &0\farcs8	&B\\
\noalign{\smallskip}\hline
\end{tabular}
\end{table*}

\subsection{Redshifts and magnitudes}

\begin{table}
\caption{Detected emission lines and redshifts of QSOs A--C.}
\label{he2347_redshifts}
\centering\scriptsize
\begin{tabular}{llll}
\hline\hline\noalign{\smallskip}
Object 			&Emission line 		&$\lambda_\mathrm{obs}~$~[\AA]&$z$\\
\noalign{\smallskip}\hline\noalign{\smallskip}
QSO~J23503$-$4328	&Ly$\alpha$		&$3989\pm4$	&$2.281\pm0.003$\\
                        &\ion{N}{v}		&$4070\pm8$	&$2.282\pm0.006$\\
                        &\ion{Si}{iv}+\ion{O}{iv}]&$4585\pm8$	&$2.276\pm0.006$\\
			&\ion{C}{iv}		&$5082\pm4$	&$2.281\pm0.003$\\
			&\ion{C}{iii}]		&$6253\pm7$	&$2.276\pm0.004$\\
			&\ion{Mg}{ii}		&$9196\pm12$	&$2.286\pm0.004$\\\cline{4-4}
			&			&		&$2.282\pm0.003$\\
QSO~J23500$-$4319	&\ion{Si}{iv}+\ion{O}{iv}]&$4613\pm6$	&$2.296\pm0.004$\\
			&\ion{C}{iv}		&$5115\pm3$	&$2.302\pm0.002$\\
			&\ion{C}{iii}]		&$6305\pm2$	&$2.303\pm0.001$\\\cline{4-4}
			&			&		&$2.302\pm0.002$\\
QSO~J23495$-$4338	&Ly$\alpha$		&$4513\pm10$	&$2.712\pm0.008$\\
			&\ion{O}{i}+\ion{Si}{ii}&$4823\pm10$	&$2.694\pm0.008$\\
			&\ion{C}{ii}		&$4930\pm10$	&$2.692\pm0.007$\\
                        &\ion{Si}{iv}+\ion{O}{iv}]&$5135\pm15$	&$2.669\pm0.011$\\
			&\ion{C}{iv}		&$5691\pm10$	&$2.674\pm0.006$\\
			&\ion{C}{iii}]		&$7028\pm10$	&$2.682\pm0.005$\\\cline{4-4}
			&			&		&$2.690\pm0.006$\\
\noalign{\smallskip}\hline
\end{tabular}
\end{table}

Redshifts of the 14 quasars were determined by taking every detectable emission
line into account. Line peaks were measured by eye and errors were estimated
taking into account the $S/N$ of the lines, line asymmetries and the presence of
absorption systems. The quasar redshifts were derived by weighting the
measurements of detected lines. Since high-ionisation lines suffer from
systematic blueshifts with respect to the systemic redshift 
\citep{gaskell82,tytler92,mcintosh99}, a higher weight was given to 
low-ionisation lines. Obviously blueshifted lines were discarded. Redshift
errors were estimated from the redshift differences of the remaining lines and
their estimated errors.

\begin{figure}
\resizebox{\hsize}{!}{\includegraphics{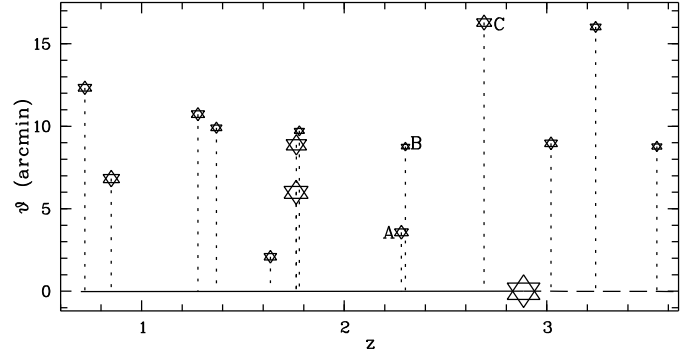}}
\caption{\label{he2347fieldplot}Distribution of separation angles $\vartheta$
vs. redshift $z$ of the quasars from Table~\ref{observinglog} with respect to
HE~2347$-$4342. Symbol size indicates apparent optical magnitude.}
\end{figure}

The 14 discovered quasars lie in the broad redshift range $0.720\le z\le 3.542$.
Fig.~\ref{he2347fieldplot} shows their angular separations with respect
to HE~2347$-$4342.
We find three background quasars to HE~2347$-$4342 and we identify
a pair of bright quasars at $z\simeq 1.763$ separated by $7\farcm 8$.
Three foreground quasars
(labelled A--C in Table~\ref{observinglog} and Fig.~\ref{he2347fieldplot}) are
located in the redshift range to study the transverse proximity effect. 
Table~\ref{he2347_redshifts} provides the redshift measurements for the detected
emission lines in their spectra. The redshift of QSO~J23503$-$4328 was based on
Ly$\alpha$ and \ion{C}{iv}. The measurement of the \ion{Mg}{ii} is uncertain
because of the decline of the resolving power of the 300V grism towards the red,
but yields a slightly higher redshift than the adopted one. For QSO~J23500$-$4319
we measured a consistent redshift from the \ion{C}{iv} and the \ion{C}{iii}]
line. The redshift measurement of QSO~J23495$-$4338 was difficult due to several
metal absorption line systems of which only two \ion{Mg}{ii} systems at $z=0.921$
and $z=1.518$ could be identified. In particular, \ion{Fe}{ii} absorption from
the $z=0.912$ system hampered a redshift measurement of the Ly$\alpha$ line. The
\ion{C}{iv} and the \ion{C}{iii}] lines show unidentified absorption features.
Thus, the redshift of QSO~J23495$-$4338 is heavily weighted towards the very
noisy low-ionisation lines \ion{O}{i}+\ion{Si}{ii} and \ion{C}{ii}. However,
redshift uncertainties of the foreground quasars do not significantly affect our
results.

Apparent magnitudes were derived from target aquisition images photometrically
calibrated against the standard star fields PG~2213$-$006 or Mark~A
\citep{landolt92}. Unfortunately the aquisition exposures of the faintest
quasars were too short to determine their magnitudes accurately. Magnitudes
derived from integration of the spectra are consistent with the photometric ones
after correcting for slit losses.

We note that QSO~J23507$-$4326 is variable. This quasar has been detected in
both slitless surveys and had $V\simeq 20.3$ in October 2001, $V\simeq 20.7$ in
October 2002 and $V\simeq 21.0$ in November 2004. We were able to discover this
quasar in its bright phase while missing the slightly fainter quasar
QSO~J23515$-$4324 detected only in the survey by P.~Jakobsen.

\subsection{Optical spectra of HE~2347$-$4342}

From the ESO Science Archive we retrieved the optical spectra of HE~2347$-$4342
taken with UVES at VLT UT2/Kueyen in the Large Programme
``The Cosmic Evolution of the Intergalactic Medium'' \citep{bergeron04}. Data
reduction was performed using the UVES pipeline provided by ESO
\citep{ballester00}. The vacuum-barycentric corrected co-added spectra yield a
$S/N\sim 100$ in the Ly$\alpha$ forest at $R\sim45000$. The spectrum was
normalised in the covered wavelength range
$3000\la\lambda\la 10000\,\mathrm{\AA}$ using a cubic spline interpolation
algorithm.

\subsection{Far-UV spectra of HE~2347$-$4342}

HE~2347$-$4342 is one of the two high-redshift quasars observed successfully in
the \ion{He}{ii} Ly$\alpha$ forest below $303.7822$~\AA~rest frame wavelength
with the Far Ultraviolet Spectroscopic Explorer (FUSE) at a resolution of
$R\sim 20000$, although at a $S/N\la 5$ \citep{kriss01,zheng04}. G.~Kriss and
W.~Zheng kindly provided the reduced FUSE spectrum of HE~2347$-$4342 described
in \citet{zheng04}. We adopted their flux normalisation with a power law
$f_\lambda=3.3\times 10^{-15}\left(\lambda/1000\mathrm{\AA}\right)^{-2.4}\,
\mathrm{erg}\,\mathrm{cm}^{-2}\,\mathrm{s}^{-1}\,\mathrm{\AA}^{-1}$ reddened by
the \citet{cardelli89} extinction curve assuming $E(B-V)=0.014$
\citep{schlegel98}.

\section{The Ly$\alpha$ forest near the foreground quasars}
\label{trprox_h1}

\begin{figure*}
\centering
\includegraphics{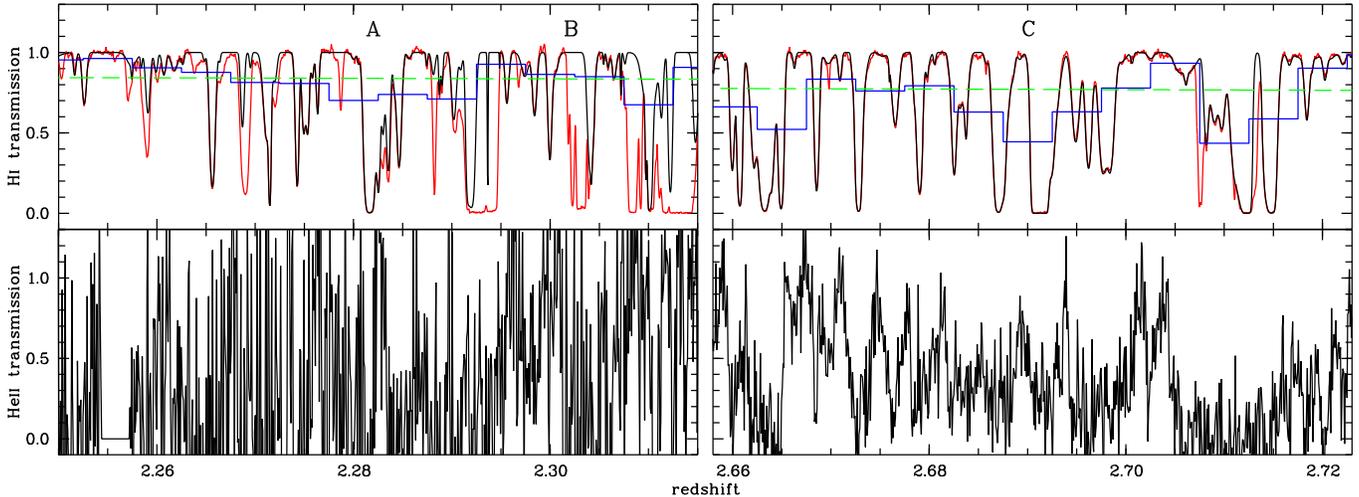}
\caption{\label{he2347_forestqso} The Ly$\alpha$ forest of HE~2347$-$4342 in the
vicinity of the foreground quasars A--C from Table~\ref{observinglog}. The upper
panels show the normalised optical spectrum of HE~2347$-$4342 including
Ly$\beta$ and metal lines (red) and the \ion{H}{i} Ly$\alpha$ transmission
obtained from the line list by T.-S.~Kim (black). The binned blue line shows the
mean \ion{H}{i} Ly$\alpha$ transmission in $\Delta z=0.005$ bins towards
HE~2347$-$4342, whereas the dashed green line indicates the expected mean
transmission $\left<T\right>^\mathrm{exp}$. The lower panels display the
corresponding \ion{He}{ii} transmission from the FUSE spectrum. [See the online
edition of the Journal for a colour version of this figure.]} 
\end{figure*}

Aiming to detect the transverse proximity effect as an underdensity ('void') in
the Ly$\alpha$ forest towards HE~2347$-$4342 we examined the forest regions in
the projected vicinity of the three foreground quasars labelled A--C in
Table~\ref{observinglog}. The \ion{H}{i} forest of HE~2347$-$4342 has been
analysed in several studies, e.g. by \citet{zheng04} and \citet{fechner07},
hereafter called Z04 and FR07, respectively. Since the line list from FR07 is
limited to $z>2.29$, T.-S.~Kim (priv.~comm.) kindly provided an independent line
list including the lower redshift Ly$\alpha$ forest ($z>1.79$). Both line lists
agree very well in their overlapping redshift range $2.29<z<2.89$.

Figure~\ref{he2347_forestqso} displays the \ion{H}{i} and the \ion{He}{ii}
forest regions near the foreground quasars A--C. The \ion{H}{i} Ly$\alpha$
forest is contaminated by metals. In particular at $z<2.332$ there is severe
contamination due to the \ion{O}{vi} absorption of the associated system of
HE~2347$-$4342 \citep{fechner04}. Because the strong \ion{O}{vi} absorption
overlaps with the projected positions of QSO~A and QSO~B it is very difficult to
obtain a well-determined \ion{H}{i} line sample in this region. Furthermore,
there is Ly$\beta$ absorption of \ion{H}{i} and \ion{He}{ii} at $z<2.294$. We
also overplot in Fig.~\ref{he2347_forestqso} the mean \ion{H}{i} Ly$\alpha$
transmission in $\Delta z=0.005$ bins obtained from T.-S.~Kim's line list and
the generally expected mean transmission over several lines of sight
$\left<T\right>^\mathrm{exp}=\mathrm{e}^{-\tau_\mathrm{eff}^\mathrm{exp}}$ with
$\tau_\mathrm{eff}^\mathrm{exp}=0.0032\left(1+z\right)^{3.37}$ \citep{kim02}.

We do not detect a significant void near the three foreground quasars, neither
in the \ion{H}{i} forest nor in the \ion{He}{ii} forest. In the vicinity of
QSO~A and QSO~B, even a careful decontamination of the optical spectrum does not
reveal a significant \ion{H}{i} underdensity. Instead, the transmission is
fluctuating around the mean. Due to the poor quality of the FUSE data in this
region ($S/N\la 2$) and the \ion{He}{ii} Ly$\beta$ absorption from higher
redshifts, a simple search for \ion{He}{ii} voids near QSO~A and QSO~B is
impossible. In the vicinity of QSO~C the \ion{H}{i} Ly$\alpha$ absorption is
slightly higher than on average. There is a small void at $z\simeq 2.702$ that
can be identified in the forests of both species. The probability of chance
occurrence of such small underdensities is high, so linking this void to QSO~C
seems unjustified. However, note that the \ion{He}{ii} absorption in the
vicinity of QSO~C ($z\sim 2.69$) is lower than at $z\sim 2.71$ in spite of the
same \ion{H}{i} absorption. This points to fluctuations in the spectral shape of
the ionising radiation near the quasar (Sect.~\ref{trprox_hardness}).

Given the luminosities and distances of our foreground quasars to the sightline
of HE~2347$-$4342, could we expect to detect the transverse proximity effect as
voids in the \ion{H}{i} forest? As in Paper~I, we modelled the impact of the
foreground quasars on the line of sight towards HE~2347$-$4342 with the parameter
\begin{equation}\label{eq_omega}
\omega\left(z\right)=\sum_{j=1}^n\frac{f_{\nu_\mathrm{LL},j}}{4\pi J_\nu\left(z\right)}
      \frac{\left(1+z_j^\prime\right)^{-\alpha_j+1}}{\left(1+z_j\right)}
      \left(\frac{\alpha_{J_\nu}+3}{\alpha_j+3}\right)
      \left(\frac{d_{L}\left(z_j,0\right)}{d_{L}\left(z_j,z\right)}\right)^2
\end{equation}
which is the ratio between the summed photoionisation rates of $n$ quasars
at redshifts $z_j$ with rest frame Lyman limit fluxes $f_{\nu_\mathrm{LL},j}$,
penetrating the absorber at redshift $z$ and the overall UV background with 
Lyman limit intensity $J_\nu$.
$d_{L}(z_j,0)$ is the luminosity distance of QSO $j$, and 
$d_{L}(z_j,z)$ is its luminosity distance as seen at the absorber; 
the redshift of the quasar as seen at the absorber is $z_j^{\prime}$
\citep{liske00}. A value $\omega\gg 1$ predicts a highly significant proximity
effect.

We assumed a constant UV background at 1~ryd of $J_\nu = 7\times
10^{-22}$~erg~cm$^{-2}$~s$^{-1}$~Hz$^{-1}$~sr$^{-1}$ \citep{scott00} with a
power-law shape $J_\nu\propto \nu^{-\alpha_{J_\nu}}$ and $\alpha_{J_\nu}=1.8$.
The quasar Lyman limit fluxes were estimated from the spectra by fitting a power
law $f_{\nu}\propto \nu^{-\alpha}$ to the quasar continuum redward of the
Ly$\alpha$ emission line, excluding the emission lines. The spectra were scaled
to yield the measured photometric magnitudes. Table~\ref{lylimitfluxes} lists
the resulting spectral indices, the \ion{H}{i} Lyman limit fluxes, and the
transverse distances.

The combined effects of QSOs~A and B result in a peak
$\omega_\mathrm{max}\simeq 0.89$, while QSO~C yields
$\omega_\mathrm{max}\simeq 0.11$. So we expect only a weak signature of the
transverse proximity effect that can be easily diluted by small-scale
transmission fluctuations around $\left<T\right>^\mathrm{exp}$. Thus, the
apparent lack of a transverse proximity effect in the \ion{H}{i} forest is no
surprise.

We can also roughly estimate the amplitude of the proximity effect in the
\ion{He}{ii} forest. Extrapolating the power laws (QSOs and background) above
4~ryd at $\eta=50$ \citep[][hereafter HM96]{haardt96} we get
$\omega_\mathrm{max}\simeq 20$ near QSO~A and $\omega_\mathrm{max}\simeq 2$ near
QSO~C. A softer background would result in higher values of $\omega$, whereas
absorption of ionising photons in the \ion{He}{ii} forest would decrease
$\omega$. However, due to the arising \ion{He}{ii} Ly$\beta$ forest and the low
$S/N$ in the FUSE data near QSOs~A and B, even high $\omega$ values do not
necessarily result in a visible \ion{He}{ii} void. In the direct vicinity of
QSO~C the \ion{He}{ii} data is not saturated, but shows no clear void structure
either. We will show in the following sections that the spectral shape of the
radiation field is a more sensitive indicator of the transverse proximity effect
than the detection of voids in the forests.

\begin{table}
\caption{Rest frame Lyman limit fluxes of foreground QSOs. A power law
$f_{\nu}\propto\nu^{-\alpha}$ is fitted to the QSO continua and
$f_{\nu_\mathrm{LL}}$ is the extrapolated \ion{H}{i} Lyman limit flux in the QSO
rest frame. $d_\perp(z)$ denotes the transverse proper distance to the line of
sight towards HE~2347$-$4342.}
\label{lylimitfluxes}
\centering\scriptsize
\begin{tabular}{llllll}
\hline\hline \noalign{\smallskip} 
QSO & Abbr.	& $z$
    & $\alpha$ 	& $f_{\nu_\mathrm{LL}}$ [$\mu$Jy]
    &$d_\perp(z)$ [Mpc]\\ 
\noalign{\smallskip} \hline \noalign{\smallskip}
QSO~J23503$-$4328 &A	&$2.282$	&$0.21$	&$16$	&$1.76$\\
QSO~J23500$-$4319 &B	&$2.302$	&$0.84$	&$1$	&$4.33$\\
QSO~J23495$-$4338 &C	&$2.690$	&$0.24$	&$29$	&$7.75$\\
\noalign{\smallskip} \hline
\end{tabular}
\end{table}

\section{The fluctuating shape of the UV radiation field towards HE~2347$-$4342}
\label{los_hardness}
\subsection{Diagnostics}

If both hydrogen and helium are highly ionised in the IGM with roughly
primordial abundances, the column density ratio
$\eta=N_\ion{He}{ii}/N_\ion{H}{i}$ indicates the softness of the UV radiation
field impinging on the absorbers. Theoretically, $\eta$ can be derived
numerically via photoionisation models of the IGM with an adopted population of
ionising sources. At the redshifts of interest, $50\la\eta\la 100$ is predicted
for a UV background generated by quasars \citep[HM96;][]{fardal98}, whereas
higher values indicate a contribution of star-forming galaxies
\citep[e.g.][hereafter HM01]{haardt01}.

The \ion{He}{ii} forest has been resolved with FUSE towards HE~2347$-$4342 and
HS~1700$+$6416, allowing a direct estimation of $\eta$ by fitting the absorption
lines \citep[][FR07]{kriss01,zheng04,fechner06}. Due to the low $S/N$ and the
strong line blending in the \ion{He}{ii} forest the \ion{He}{ii} lines have to
be fitted with absorber redshifts and Doppler parameters fixed from the fitting
of the \ion{H}{i} data of much higher quality. Generally, pure non-thermal line
broadening ($b_\ion{He}{ii}=b_\ion{H}{i}$) is assumed (however, see FR07 and
Sect.~\ref{discussion} below). The \ion{He}{ii} forest towards HE~2347$-$4342
was fitted independently by Z04 and FR07. In the following, we rely on the line
fitting results from FR07, which at any rate are consistent with those obtained
by Z04 in the redshift ranges near the quasars.

\begin{figure*}
\sidecaption
\includegraphics[width=12cm]{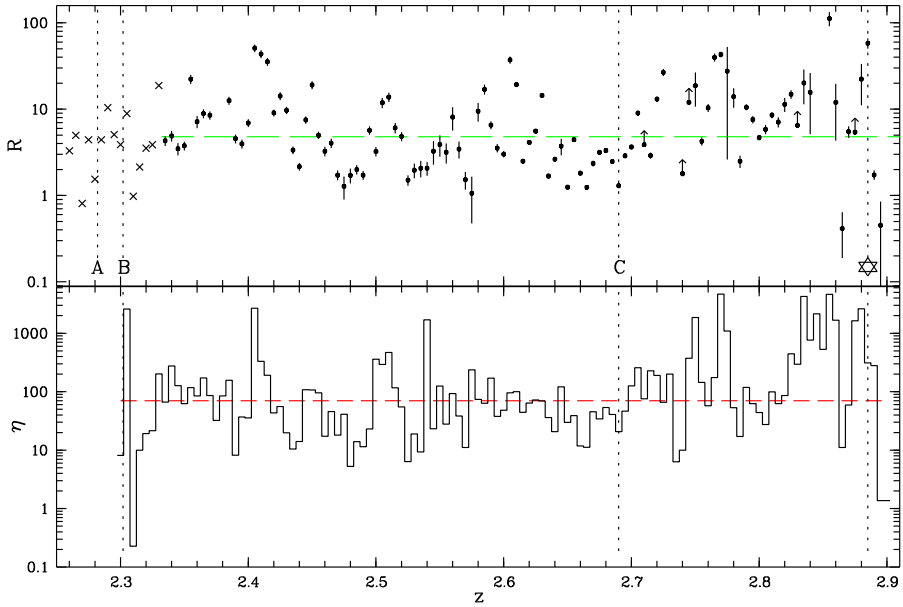}
\caption{\label{he2347forestratio} The fluctuating spectral shape of the UV
background towards HE~2347$-$4342. The upper panel shows the ratio of effective
optical depths $R$ vs.\ redshift $z$ in $\Delta z=0.005$ bins. Data points at
$z<2.332$ (crosses) have been decontaminated from \ion{O}{vi} and Ly$\beta$
absorption (see text). Foreground quasars are marked with letters and vertical
dotted lines as well as HE~2347$-$4342 (star symbol). The green dashed line
indicates the median $R\simeq 4.8$ obtained at $z>2.332$ in uncontaminated bins.
The lower panel shows the median $\eta$ from FR07 in the same redshift bins. The
red dashed line indicates the median $\eta\simeq 70$ of the line sample.}
\end{figure*}

All current studies indicate that $\eta$ is strongly fluctuating on very small
scales in the range $1\la\eta\la 1000$. The median column density ratio towards
HE~2347$-$4342 is $\eta\simeq 62$ (Z04), whereas \citet{fechner06} find a higher
value of $\eta\simeq 85$ towards HS~1700$+$6416. Both studies find evidence for
an evolution of $\eta$ towards smaller values at lower redshifts. However, only
part of the scatter in $\eta$ is due to redshift evolution and statistical
errors, so the spectral shape of the UV radiation field has to fluctuate (FR07).
Although the analyses of both available lines of sight give consistent results,
cosmic variance may bias the derived median $\eta$ and its evolution. This is of
particular interest for our study, since we want to reveal local excesses of low
$\eta$ near the quasars with respect to the median
(Sect.~\ref{trprox_hardness}). Clearly, more lines of sight with \ion{He}{ii}
absorption would be required to yield tighter constraints on the redshift
evolution of $\eta$.

The detailed results of visual line fitting may be subjective and may depend on
the used fitting software. In particular, ambiguities in the decomposition of
blended \ion{H}{i} lines can affect the derived $\eta$ values \citep{fechner07b}.
Therefore we also analyse the UV spectral shape variations using the ratio of
the effective optical depths
\begin{equation}
R=\frac{\tau_{\mathrm{eff},\ion{He}{ii}}}{\tau_{\mathrm{eff},\ion{H}{i}}}.
\end{equation}
As introduced in Paper~I, this parameter is a resolution-independent estimator
of the spectral shape of the UV radiation field with small (high) $R$ values
indicating hard (soft) radiation on a certain redshift scale $\Delta z$.
\citet{shull04} followed a similar approach by taking
$\eta\simeq 4\tau_\ion{He}{ii}/\tau_\ion{H}{i}$ for a restricted $\tau$ range on
scales of $\Delta z=1.6\times 10^{-4}$ and $\Delta z=6.6\times 10^{-4}$.
However, this scaling relation between $\tau$ and $\eta$ is only valid at the
centre of an absorption line \citep{miralda-escude93}. The column density ratio
is defined per absorption line and not as a continuous quantity, whereas $R$ can
be defined on any scale. While $R$ and $\eta$ are correlated (see below), there
is no simple conversion between $R$ and $\eta$ and the correlation will depend
on the adopted redshift scale of $R$.

\subsection{Fluctuations in $R$ and $\eta$ along the line of sight}
\label{forestratio_los}

We obtained $R(z)$ by binning both normalised Ly$\alpha$ forest spectra of
\ion{H}{i} and \ion{He}{ii} into aligned redshift bins of $\Delta z=0.005$ in
the range $2.3325<z<2.8975$ and computed
$R=\ln{\left<T_\ion{He}{ii}\right>}/\ln{\left<T_\ion{H}{i}\right>}$ with the
mean transmission $\left<T_\ion{He}{ii}\right>$ and $\left<T_\ion{H}{i}\right>$.
The choice of the redshift binning scale was motivated by the typical scale of
$\eta$ fluctuations $0.001\la \Delta z\la 0.03$ \citep[][FR07]{kriss01}. We
adopted the binning procedure by \citet{telfer02} in order to deal with original
flux bins that only partly overlap with the new bins. The errors were computed
accordingly. Due to the high absorption and the low $S/N$ of the \ion{He}{ii}
data we occasionally encountered unphysical values
$\left<T_\ion{He}{ii}\right>\le 0$. These were replaced by their errors,
yielding lower limits on $R$. We mostly neglected the usually small metal
contamination in the computation of $\left<T_\ion{H}{i}\right>$ in the
Ly$\alpha$ forest because the errors in $R$ are dominated by the low $S/N$
and the more uncertain continuum level of the \ion{He}{ii} spectrum. The FUSE
data in the redshift bins at $z=2.375$, $2.380$, $2.730$, $2.735$, $2.845$ and
$2.850$ are contaminated by galactic $\mathrm{H}_2$ absorption, so no $R$
measurement on the full scale of $\Delta z=0.005$ can be performed there.

At $2.29\la z_\mathrm{Ly\alpha}\la 2.33$ the \ion{H}{i} Ly$\alpha$ forest is
severely contaminated by \ion{O}{vi} from the associated system of
HE~2347$-$4342 \citep{fechner04}. Furthermore, the Ly$\beta$ forest of both
species emerges at $z_\mathrm{Ly\alpha}<2.294$. Because this excess absorption
would bias the direct estimation of $R$ in the spectra, we tried to
decontaminate the forests at $z<2.332$. $\left<T_\ion{H}{i}\right>$ was
computed from the \ion{H}{i} Ly$\alpha$ forest reconstructed from the line list
by T.-S.~Kim (Sect.~\ref{trprox_h1}). The corresponding
$\left<T_\ion{He}{ii}\right>$ was obtained after dividing the FUSE data by the
simulated Ly$\beta$ absorption of the lines at higher redshift. Since the
decontamination depends on the validity of the \ion{He}{ii} line parameters as
well as on the completeness of the \ion{H}{i} line list in the complex region
contaminated by \ion{O}{vi}, the derived $R$ values at $z<2.332$ have to be
regarded as rough estimates.

The resulting $R(z)$ is shown in the upper panel of
Fig.~\ref{he2347forestratio}. The optical depth ratio strongly fluctuates around
its median value $R\simeq 4.8$ obtained for uncontaminated redshift bins,
indicating spectral fluctuations in the UV radiation field. We also show in
Fig.~\ref{he2347forestratio} the median $\eta(z)$ on the same redshift bins
based on the line fitting results in FR07. Also the median $\eta$ strongly
fluctuates with a slight trend of an increase with redshift (Z04). Clearly, the
data is inconsistent with a spatially uniform UV background, but the median
$\eta\simeq 70$ of the line sample is consistent with quasar-dominated models of
the UV background.
A comparison of $R(z)$ and $\eta(z)$ reveals that both quantities are correlated.
The Spearman rank order correlation coefficient is $r_\mathrm{S}=0.67$ with a
probability of no correlation $P_\mathrm{S}= 6\times 10^{-15}$.

There is a scatter in the relation between $R$ and $\eta$, which is due to noise
in the \ion{He}{ii} data and due to the fact that $R$ is a spectral softness
indicator that is smoothed in redshift.
Therefore, in addition to the UV spectral shape, $R$ will depend on the density
fluctuations of the Ly$\alpha$ forest on the adopted scale. In order to estimate
the scatter in $R$ due to these density fluctuations, we simulated \ion{H}{i}
and \ion{He}{ii} Ly$\alpha$ forest spectra. We generated 100 \ion{H}{i} forests
with the same overall redshift evolution of
$\tau_{\mathrm{eff},\ion{H}{i}}^\mathrm{exp}=0.0032\left(1+z\right)^{3.37}$
\citep{kim02} based on the empirical line distribution functions in redshift $z$,
column density $N_\ion{H}{i}$ and Doppler parameter $b_\ion{H}{i}$
\citep[e.g.][]{kim01}. We modelled each forest as a composition of lines with
Voigt profiles using the approximation by \citet{teppergarcia06}. The spectral
resolution ($R\sim 42000$) and quality ($S/N\sim 100$) closely matches the
optical data of HE~2347$-$4342. The corresponding \ion{He}{ii} forests were
generated at FUSE resolution with a $S/N=4$ for four constant values of
$\eta=10,\,20,\,50$ and 100. We assumed pure non-thermal broadening of the
lines. Then we computed $R$ at $2\le z\le 3$ on our adopted scale
$\Delta z=0.005$, yielding 20000 $R$ measurements for each considered $\eta$.
For convenience we took out the general redshift dependence of
$\tau_{\mathrm{eff},\ion{H}{i}}$ via dividing by the expected effective optical
depth $\tau_{\mathrm{eff},\ion{H}{i}}^\mathrm{exp}$, so
\begin{equation}
D=\frac{\tau_{\mathrm{eff},\ion{H}{i}}}{\tau_{\mathrm{eff},\ion{H}{i}}^\mathrm{exp}}
\end{equation}
is a measure of \ion{H}{i} overdensity ($D>1$) or underdensity ($D<1$).

\begin{figure}
\resizebox{\hsize}{!}{\includegraphics{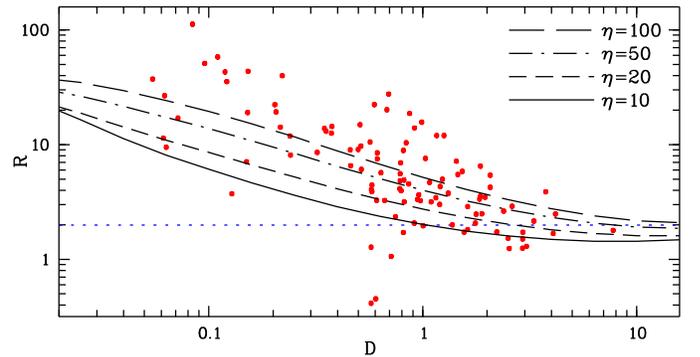}}
\caption{\label{he2347_ratiostat} Dependence of $R$ on
$D=\tau_{\mathrm{eff},\ion{H}{i}}/\tau_{\mathrm{eff},\ion{H}{i}}^\mathrm{exp}$
for different simulated values of $\eta$. The black lines indicate the
polynomial fits to the simulated distributions in logarithmic space. Red filled
circles represent the measured $R(D)$ towards HE~2347$-$4342 in uncontaminated
bins at $z>2.332$. The horizontal dotted line denotes $R=2$. [See the online
edition of the Journal for a colour version of this figure.]}
\end{figure}

In Fig.~\ref{he2347_ratiostat} we show the relation $R(D)$ obtained from the
Monte Carlo simulations and compare it to the distribution observed towards
HE~2347$-$4342. The simulated $R(D)$ can be fitted reasonably with a 3rd order
polynomial in logarithmic space, yielding a general decrease of $R$ with $D$ for
every $\eta$. The root-mean-square scatter increases from $0.13$~dex for
$\eta=10$ to $0.18$~dex for $\eta=100$. At $D\ga 3$ the $R(D)$ distribution
flattens due to saturation of high-column density absorbers on the flat part of
the curve of growth. The flattening causes substantial overlap between the
simulated $R$ distributions at $D\ga 5$, making $R$ increasingly insensitive to
the underlying $\eta$. However, at $D\la 3$ hard radiation and soft radiation
can be reasonably well distinguished. We also overplot the measured $R(D)$
towards HE~2347$-$4342 in Fig.~\ref{he2347_ratiostat}. The observed distribution
is inconsistent with a constant $\eta$, but the majority of values falls into
the modelled $\eta$ range. While many high $R$ values indicate $\eta>100$,
values with $R\la 2$ correspond to $\eta\la 20$ at $D\la 3$. Thus, the very low
$R$ values always indicate a hard radiation field up to moderate overdensities.
As we will see in the next section, the saturation effect probably does not play
a role in relating a hard radiation field to the nearby quasars.

\subsection{The UV radiation field near the quasars}
\label{trprox_hardness}

We now investigate in greater detail the spectral shape of the UV radiation
field near the four quasars with available data on $R$ and $\eta$: the
background quasar HE~2347$-$4342 and the foreground QSOs A--C. Due to the small
number of comparison values derived from only two lines of sight, we will adopt
$\eta=100$ as a characteristic value for the overall UV background at $z>2.6$
(HE~2347$-$4342, QSO~C) and a value of $\eta=50$ at $z\sim 2.3$ (QSOs A and B).
The former value is close to the median $\eta=102$ obtained by \citet{fechner06}
at $2.58<z<2.75$ towards HS~1700$+$6416, whereas the latter $\eta$ value
accounts for the probable evolution of $\eta$ with redshift. Furthermore, we
will compare the $\eta$ values in the vicinity of the quasars to models of the
UV background.

\subsubsection{HE~2347$-$4342}
\label{hardness_he2347}

A close inspection of Fig.~\ref{he2347forestratio} reveals a strongly
fluctuating radiation field near HE~2347$-$4342 with some very small, but also
high $R$ values.
Also the column density ratio shows large fluctuations ($1\la\eta\la 1000$) with
six $\eta\la 10$ absorbers out of the 20 absorbers at $z>2.86$. These strong
variations of the spectral shape are likely due to radiative
transfer effects in the associated absorption system causing an apparent
lack of the proximity effect of HE~2347$-$4342 \citep{reimers97}. The high
\ion{He}{ii} column densities of the associated system may soften the quasar
radiation with increasing distance and Fig.~\ref{he2347forestratio} supports
this interpretation. Due to the probable strong softening of the
hard quasar radiation on small scales, the relative spectral hardness near
HE~2347$-$4342 is only revealed by individual low $\eta$ values instead of robust
median values. However, also the highly ionised metal species of the
associated system \citep{fechner04} favour the presence of hard QSO radiation.
Thus we conclude that despite the lack of a radiation-induced void near
HE~2347$-$4342, its impact onto the IGM can be detected via the relative spectral
hardness of the UV radiation field. The three $R<2$ values near HE~2347$-$4342
have $D<3$, so they are probably not affected by saturation.

\subsubsection{QSOs~A and B}

If our decontamination of the Ly$\alpha$ forests near the two $z\sim 2.3$ QSOs
A and B is correct, $R$ should reflect UV spectral
shape variations also in that region. Indeed,
the redshift bin at $z=2.280$ next to QSO~A ($z=2.282$) is a local $R$ minimum
with $R\simeq 1.5$. At $z=2.270$ we find $R\simeq 0.8$. At the redshift of QSO~B
($z=2.302$) the radiation field is quite soft, but we note a low $R\sim 1$ at
$z=2.310$. We obtain $D<3$ for the four $R\la 2$ values near QSO~A and QSO~B, so
saturation is not relevant, and the low $R$ values correspond to low $\eta$ values.

\begin{figure}
\resizebox{\hsize}{!}{\includegraphics[angle=270]{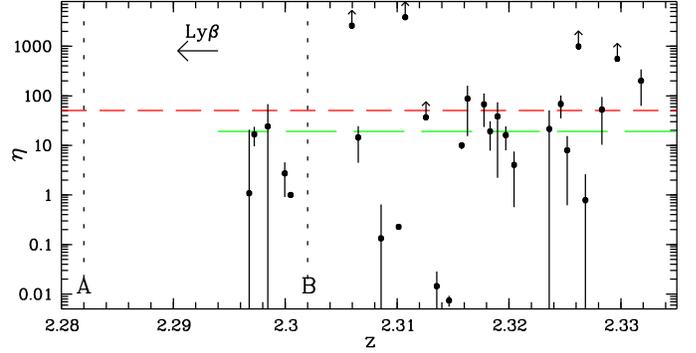}}
\caption{\label{he2347etaplot_ab} Column density ratio $\eta$ vs. redshift $z$
in the vicinity of QSO~A and QSO~B. The long (short) dashed line indicates the
median $\eta\simeq 19$ in this redshift range ($\eta=50$ for a UV background
generated by quasars). At $z<2.294$ the \ion{He}{ii} Ly$\beta$ forest sets in.}
\end{figure}

The measured $\eta$ values in this redshift region are presented in
Fig.~\ref{he2347etaplot_ab}. The error bars are only indicative, since blended
line components are not independent and the \ion{He}{ii} column densities are
derived with constraints from the \ion{H}{i} forest. Lower limits on $\eta$
result from features detected in \ion{He}{ii} but not in \ion{H}{i}. Due to
ambiguities in the line profile decomposition at the \ion{H}{i} detection limit
and the present low quality of the \ion{He}{ii} data it is hard to judge
the reality of most of these added components \citep{fechner07b}. Nevertheless,
since $\eta$ for adjacent lines may be not independent due to line blending, we
must include the lower limits in the analysis. At $z<2.294$ the fitting of
\ion{He}{ii} lines becomes unreliable due to the arising Ly$\beta$ forest.
Therefore, no direct estimates of $\eta$ can be obtained in the immediate
vicinity of QSO~A. Furthermore, the \ion{H}{i} line sample may be incomplete or
the line parameters may be not well constrained due to blending with the
\ion{O}{vi} of the associated system of HE~2347$-$4342.

Considering these caveats, the median $\eta\simeq 19$ obtained for the values at
$z<2.332$ shown in Fig.~\ref{he2347etaplot_ab} is only an estimate. Nevertheless,
this is much lower than the typical values $\eta\sim 50$ found at $z\sim 2.3$
towards HS~1700$+$6416 \citep{fechner06}. Moreover, it is also lower than at
slightly higher redshifts towards HE~2347$-$4342. For
instance, the median $\eta$ increases to $\eta=79$ in the redshift range
$2.35\le z\le 2.40$. This is inconsistent with the smooth
redshift evolution of $\eta$ on large scales inferred by Z04 and
\citet{fechner06} for both available sightlines. Thus, we infer an excess of
hard radiation in the vicinity of QSO~A and QSO~B. The most extreme $\eta$
values are located in the projected vicinity of QSO~B, with 6 lines reaching
$\eta<1$. If estimated correctly, these low $\eta$ values require local hard
sources and cannot be generated by the diffuse UV background. Both foreground
quasars could be responsible for the hard radiation field because of similar
light travel times to the probably affected absorbers
($t_\mathrm{A}\simeq 2t_\mathrm{B}$).

\begin{figure*}
\sidecaption
\includegraphics[width=12cm]{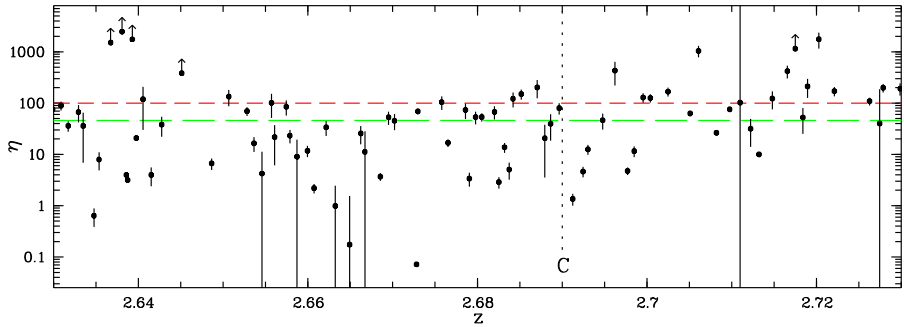}
\caption{\label{he2347etaplot_c} Column density ratio $\eta$ vs. redshift $z$ in
the vicinity of QSO~C. The short dashed line denotes $\eta=100$ that is consistent
with the median $\eta\simeq 102$ obtained in the range $2.58<z<2.75$ towards
HS~1700$+$6416 \citep{fechner06}. The long dashed line indicates the median
$\eta\simeq 46$ obtained for the shown $\eta$ values ($2.63<z<2.73$).}
\end{figure*}

\subsubsection{QSO~C}
\label{hardness_qsoc}

Since metal contamination of the \ion{H}{i} forest is small in the projected
vicinity of QSO~C (Fig.~\ref{he2347_forestqso}), the UV spectral shape is better
constrained here than near QSO~A and QSO~B.
From Fig.~\ref{he2347forestratio} we note a local $R$ minimum ($R\simeq 1.3$)
that exactly coincides with the redshift of QSO~C ($z=2.690$). At higher redshifts
$R$ rises, possibly indicating a softer ionising field. However, at $2.63\la z\la 2.695$
the optical depth ratio is continuously below the median with $R<2$ in five
redshift bins. Due to the \ion{H}{i} overdensities near QSO~C, all $R<2$ values
have $D>1$, but only the bin at $z=2.635$ has $D\simeq 4$, so the remaining ones
may still indicate low column density ratios $\eta$.

Figure~\ref{he2347etaplot_c} displays the $\eta$ values from FR07 in the
redshift range $2.63<z<2.73$ in the projected vicinity of QSO~C. For comparison,
we also indicate $\eta=100$ that is consistent with the median $\eta=102$
towards HS~1700$+$6416 in this redshift range \citep{fechner06}. While the data
generally shows strong fluctuations around the median over the whole covered
redshift range \citep[Z04;][]{fechner06}, there is an apparent excess of small
$\eta$ values near QSO~C indicating a predominantly hard radiation field. From
the data, a median $\eta\simeq 46$ is obtained at $2.63<z<2.73$ including the
lower limits on $\eta$. The median $\eta$ near QSO~C is lower than the median
$\eta$ towards HS~1700$+$6416 by a factor of two and also slightly lower than
the $\eta$ obtained for spatially uniform UV backgrounds generated by quasars.
The relative agreement of the median $\eta$ near QSO~C and hard versions of
quasar UV background models may result from the softening of the quasar
radiation by the IGM at the large proper distances $d\ga 7.75$~Mpc considered
here (Table~\ref{lylimitfluxes}). This will be further explored in
Sect.~\ref{etamodel}. The larger contrast between the median $\eta$ near QSO~C
and the median $\eta$ towards HS~1700$+$6416 yields stronger evidence for a
local hardening of the UV radiation near QSO~C. However, this comparison value
derived from the single additional line of sight tracing this redshift range
may be biased itself.

Near QSO~C the column density ratio still fluctuates and is not homogeneously
low as naively expected. We also note an apparent offset of the low $\eta$
region near QSO~C towards lower redshift due to fewer absorbers with low $\eta$
at $z>2.69$. While some of the fluctuations can be explained by uncertainties
to recover $\eta$ reliably from the present data, the very low $\eta\le 10$
values ($\simeq 24$\% of the data in Fig.~\ref{he2347etaplot_c}) are likely
intrinsically low. These $\eta$ values are in conflict with a homogeneous
diffuse UV background, and are likely affected by a local hard source. In
Sect.~\ref{uncertainties} we will estimate the error budget of $\eta$ by Monte
Carlo simulations.

If QSO~C creates a fluctuation in the spectral shape of the UV background, the
distance between the quasar and the line of sight implies a light travel time of
$t= 25$~Myr. The low (high) redshift end of the region shown in
Fig.~\ref{he2347etaplot_c} corresponds to a light travel time of $64$~Myr
($44$~Myr). Since these light travel times are comparable, we argue that it is
important to consider not only the immediate projected vicinity of QSO~C to be
affected by the proximity effect (see also Fig.~\ref{he2347_isotime} below).

In summary, both spectral shape indicators $R$ and $\eta$ indicate a
predominantly hard UV radiation field near all four known quasars in this field.
Many $\eta$ values in the projected vicinity of the quasars indicate a harder
radiation than expected even for model UV backgrounds of quasars alone. This
points to a transverse proximity effect detectable via the relative spectral
hardness. However, there are other locations along the line of sight with an
inferred hard radiation field, but without an associated quasar, most notably
the regions at $z\sim 2.48$ and $z\sim 2.53$ (Fig.~\ref{he2347forestratio}).
Before discussing these in detail (Sect.~\ref{hardrad_noqso}), we search for
additional evidence for hard radiation near the foreground quasars by analysing
nearby metal line systems.

\section{Constraints from metal line systems}
\label{metalsystem}

Observed metal line systems provide an additional
tool to constrain the spectral shape of the ionising radiation. Since
photoionisation modelling depends on several free parameters, appropriate
systems should preferably show many different ionic species. \citet{fechner04}
analysed the associated metal line system of HE~2347$-$4342 and found evidence
for a hard quasar spectral energy distribution at the absorbers with highest
velocities that are probably closest to the quasar. Their large \ion{He}{ii}
column densities probably shield the other absorbers which are better modelled
with a softer radiation field. The results by \citet{fechner04} are consistent
with the more direct hardness estimators $R$ and $\eta$ near HE~2347$-$4342
(Sect.~\ref{hardness_he2347}).

In the spectrum of HE~2347$-$4342 an intervening metal line system is
detected at $z = 2.7119$ which is close to the redshift of QSO~C
($\Delta z = 0.022$) at a proper distance of $d\simeq 10.0$~Mpc.
At $z=2.2753$ there is another system showing multiple components of
\ion{C}{iv} and \ion{N}{v} as well as only weak \ion{H}{i} absorption
($N_{\ion{H}{i}} < 10^{13.7}\,\mathrm{cm}^{-2}$). The presence of \ion{N}{v}
and weak \ion{H}{i} features with associated metal absorption are characteristic
of intrinsic absorption systems exposed to hard radiation. 
Due to the small proper distance to QSO~A ($d\simeq 3.1$~Mpc) this system is
probably illuminated by the radiation of the close-by quasar.
A third suitable metal line system at $z=2.3132$ is closer to QSO~B
($d\simeq 6.1$~Mpc) than to QSO~A ($d\simeq 12.0$~Mpc). But since QSO~A is
much brighter than QSO~B (Table~\ref{lylimitfluxes}), the metal line system
at $z=2.3132$ might be affected by both quasars.
Due to their small relative velocities with respect to the quasars of
$<3000$~$\mathrm{km}\,\mathrm{s}^{-1}$ the systems are likely associated to the
quasars \citep[e.g.][]{weymann81}.

In order to construct CLOUDY models \citep[][version 05.07]{ferland98}
we assumed a single-phase medium, i.e.\ all observed ions arise from
the same gas phase, as well as a solar abundance pattern
\citep{asplund05} at a constant metallicity throughout the
system. Furthermore, we assumed pure photoionisation and neglected a
possible contribution of collisional ionisation.
The absorbers were modelled as distinct, plane-parallel slabs of
constant density testing different ionising spectra.

\begin{table}
\caption{Measured column densities of the metal line system at
$z=2.2753$. Several components of \ion{H}{i} remain unresolved.}
\label{mls_columns}
\centering\scriptsize
$$
\begin{array}{crccc}
\hline\hline\noalign{\smallskip}
\#&v\,[\mathrm{km\,s}^{-1}]&\ion{H}{i}&\ion{C}{iv}&\ion{N}{v}\\
\noalign{\smallskip}\hline\noalign{\smallskip}
1&-106.2& &13.25\pm0.59&12.70\pm0.27\\
2& -94.3&\raisebox{1.5ex}[1.5ex]{$13.634\pm0.005$}&13.26\pm0.57&12.40\pm0.42\\
3& -45.8& &12.76\pm0.39&12.43\pm0.14\\
4& -32.0&\raisebox{1.5ex}[1.5ex]{$13.319\pm0.015$}&12.80\pm0.40&12.88\pm0.06\\
5&  0.0&13.232\pm0.019&13.17\pm0.03&12.94\pm0.02\\
6& 44.8&12.604\pm0.020&12.51\pm0.09&12.21\pm0.05\\
7& 91.5&13.042\pm0.007&12.79\pm0.05&11.98\pm0.11\\
\noalign{\smallskip}\hline
\end{array}
$$
\end{table}

\subsection{The system at $z=2.275$ near QSO~A}

\begin{figure}
\resizebox{\hsize}{!}{\includegraphics[bb=34 595 372 780,clip=]{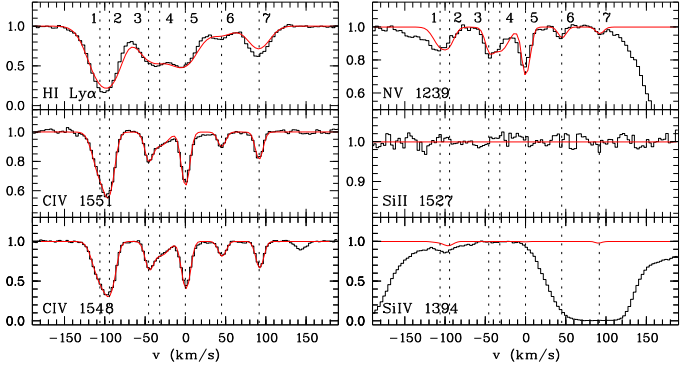}}
\caption{\label{mls_z2.2753}
Metal line system at $z = 2.2753$ towards HE~2347$-$4342. The
displayed profiles assume the rescaled HM01 background. Zero velocity
corresponds to $z = 2.2753$.}
\end{figure}

The system at $z = 2.2753$ shows seven components of \ion{C}{iv} and
\ion{N}{v} along with unsaturated features of \ion{H}{i}
(Fig.~\ref{mls_z2.2753}). The absorber densities are constrained by
the \ion{C}{iv}/\ion{N}{v} ratio. For the HM01 background scaled to yield
$\log J_{\mathrm{b}} = -21.15$ at the \ion{H}{i} Lyman limit
\citep{scott00}, we derive densities in the range 
$10^{-4.38}$ to $10^{-3.35}\,\mathrm{cm}^{-3}$ at a
metallicity of $\sim 0.6$ solar. The estimated absorber sizes are
$\sim 5\,\mathrm{kpc}$ or even smaller, where the sizes are computed
according to $N_{\element{H}} = n_{\element{H}} l$ with the absorbing
path length $l$.
With an additional contribution by QSO~A, modelled as a power
law with $\alpha = 0.21$ and \ion{H}{i} Lyman limit intensity
$\log J_{\mathrm{q}} = -21.9$ at the location of the absorber, we
obtain an even higher metallicity of $\sim 11$ times solar.
Densities in the range $10^{-2.95}$ to $10^{-2.01}\,\mathrm{cm}^{-3}$
are found leading to very small absorbers of $\lesssim 10\,\mathrm{pc}$.

Both models lead to unusually high metallicities and very small
absorber sizes. However, \citet{schaye07} recently reported on a large
population of compact high-metallicity absorbers.
Using the HM01 background they found typical sizes of $\sim
100\,\mathrm{pc}$ and densities of $10^{-3.5}\,\mathrm{cm}^{-3}$
for absorbers with nearly solar or even super-solar metallicities.
In fact, this system is part of the sample by \citet{schaye07}.

Since the system exhibits only a few different species, it is
impossible to discriminate between the soft and the hard radiation
model. In principle, both models lead to a consistent description of
the observed metal lines.
The soft HM01 UV background yields $\eta \sim 170$ for the modelled
absorbers, whereas the model including the hard radiation of QSO~A leads to 
$\eta \sim 10$. Recall that the \ion{He}{ii} forest cannot be
used to measure $\eta$ directly due to blending with Ly$\beta$ features
and very low $S/N$.

\subsection{The systems near QSO~B and QSO~C}

The systems at $z=2.3132$ and $z = 2.7119$ are located near QSO~B
and QSO~C, respectively. Only few ions are observed and some of them may even be
blended. Therefore, no significant conclusions based on CLOUDY models can be
drawn. Using the column density estimates we find that the system at $z=2.7119$
close to QSO~C exhibiting \ion{C}{iv} and \ion{O}{vi} can be described
consistently with a HM01+QSO background.
Models assuming a quasar flux of $\log J_{\mathrm{q}}\gtrsim
-22.5$ seen by the absorber yield $\eta\lesssim 40$, consistent with
the direct measurements from the \ion{He}{ii} forest. However, the metal
transitions alone do not provide strong constraints.

The system at $z=2.3132$ shows \ion{C}{iv} in six components along with
\ion{Si}{iv} and \ion{Si}{iii}. 
The Lyman series of this system suffers from severe blending
preventing a reliable estimation of the \ion{H}{i} column density.
Therefore metallicities and absorber sizes cannot be estimated.
Adopting our column density estimates we infer that this system
can be reasonably modelled with or without a specific quasar contribution.

\section{Discussion}\label{discussion}

\subsection{The transverse proximity effect in spectral hardness}
Fourteen quasars have been found in the vicinity of HE~2347$-$4342 of which three
are located in the usable part of the \ion{H}{i} Ly$\alpha$ forest towards
HE~2347$-$4342. No \ion{H}{i} underdensity is detected near these foreground
quasars even when correcting for contamination by the \ion{O}{vi} absorption from
the associated system of HE~2347$-$4342. An estimate of the predicted effect
confirms that even if existing, the classical proximity effect is probably too
weak to be detected on this line of sight due to the high UV background at 1~ryd
and small-scale variance in the \ion{H}{i} transmission (Sect.~\ref{trprox_h1}).

However, the analysis of the spectral shape of the UV radiation field near the
foreground quasars yields a markedly different result. The spectral shape is
fluctuating, but it is predominantly hard near HE~2347$-$4342 and the known
foreground quasars. Close to QSO~C, both estimators $R$ and $\eta$ are consistent
with a significantly harder radiation field than on average. There is a sharp
$R$ minimum located precisely at the redshift of the quasar, but embedded in a
broader region of low $R$ values statistically consistent with a hard radiation
field of $\eta\la 10$ (Fig.~\ref{he2347forestratio}). The column density ratio
$\eta$ is also lower than on average and indicates a harder radiation field than
obtained for quasar-dominated models of the UV background
(Fig.~\ref{he2347etaplot_c}). Because of line blending, only one of the three
metal line systems detected near the foreground quasars can be used to estimate
the shape of the ionising field. The metal line system at $z=2.275$ can be
described reasonably by the HM01 background with or without a local ionising
component by QSO~A. The \ion{He}{ii} forest does not provide independent
constraints for this absorber. Line blending prevents an unambiguous detection
of \ion{O}{vi} at $z=2.712$, leaving the shape of the ionising field poorly
constrained without taking into account the \ion{He}{ii} forest. Thus, the
systems show highly ionised metal species, but our attempts to identify a local
quasar radiation component towards them remain inconclusive.

The most probable sources for the hard radiation field at
$z\sim 2.30$ and $z\sim 2.69$ towards HE~2347$-$4342 are the nearby foreground
quasars. In particular, the absorbers with $\eta\la 10$ have to be located in
the vicinity of an AGN, since the filtering of quasar radiation over large
distances results in $\eta\ga 50$. Also star-forming galaxies close to the line
of sight cannot yield the low $\eta$ values, since they are unable to produce
significant numbers of photons at $h\nu>54.4$~eV
\citep{leitherer99,smith02,schaerer03}. We conclude that there is evidence for a
transverse proximity effect of QSO~C detectable via the relative spectral
hardness. There are also indications that QSO~A and QSO~B show the same effect,
although contamination adds uncertainty to the spectral shape variations in
their projected vicinity.

Given these incidences of a hard radiation field near the quasars, how do these
results relate to those of Paper~I, in which we investigated the line of sight
towards Q~0302$-$003? Both lines of sight show \ion{He}{ii} absorption and on
both lines of sight we find evidence for a predominantly hard radiation field
near the quasars in the background and the foreground. However, the decrease of
$\eta$ near quasars towards Q~0302$-$003 appears to be much smoother than
towards HE~2347$-$4342.

There are several reasons for the lack of small-scale spectral shape variations
on the line of sight to Q~0302$-$003. First, the low-resolution STIS spectrum of
Q~0302$-$003 does not resolve the \ion{He}{ii} lines and limits the visible
scale of fluctuations to $\Delta z\gtrsim 0.006$ (Paper~I). Much smaller scales
can be probed in the resolved \ion{He}{ii} forest of HE~2347$-$4342, but the
fitting of blended noisy \ion{He}{ii} features may result in artifical $\eta$
variations. We will discuss the uncertainties of $\eta$ below
(Sect.~\ref{uncertainties}). Second, Q~0302$-$003 ($z=3.285$) probes higher
redshifts, where the \ion{He}{ii} fraction in the IGM is significantly higher
and the inferred radiation field is very soft ($\eta\sim 350$ in the
Gunn-Peterson trough). Therefore, the impact of a hard source on the spectral
shape is likely to be more pronounced than at lower redshifts after the end of
\ion{He}{ii} reionisation, where $\eta$ of the UV background gradually decreases.

\subsection{The decrease of $\eta$ near QSO~C}
\label{etamodel}
We now investigate quantitatively whether the foreground quasars are capable of
creating a hardness fluctuation on the sightline towards HE~2347$-$4342.
Unfortunately, since only one quasar is located in an uncontaminated region of
the Ly$\alpha$ forests, we can present sufficient evidence only for QSO~C. For
the other two quasars the data is too sparse and contamination adds uncertainty
to the derived $\eta$, but in principle QSO~A should also show a strong effect,
because its Lyman limit flux penetrating the line of sight is $\sim 8$ times
higher than the one of QSO~C.

\citet{heap00} and \citet{smette02} presented simple models of the decrease of
$\eta$ in front of a quasar taking into account the absorption of ionising
photons by the IGM. In a highly photoionised IGM with helium mass fraction
$Y\simeq0.24$ and temperature $T\simeq 2\times 10^4$~K we have
\begin{equation}
\eta\simeq\frac{Y}{4\left(1-Y\right)}\frac{\alpha_\ion{He}{ii}}{\alpha_\ion{H}{i}}
\frac{\Gamma_\ion{H}{i}}{\Gamma_\ion{He}{ii}}
\simeq 0.42\frac{\Gamma_\ion{H}{i}}{\Gamma_\ion{He}{ii}},
\end{equation}
where $\Gamma_i$ and $\alpha_i$ are the photoionisation rate and the radiative
recombination coefficient for species $i$ \citep{fardal98}. The photoionisation
rate is $\Gamma_i=\Gamma_{i,\mathrm{b}}+\Gamma_{i,\mathrm{q}}$ with a
contribution of the background and the quasar. The contribution of the quasar to
the photoionisation rate of species $i$ at the $j$th absorber in front of it
($z_j>z_{j+1}$) is
\begin{eqnarray}
\Gamma_{i,\mathrm{q}}(z_j)&=&\frac{\sigma_i f_{\nu,i}}{h(1+z_\mathrm{q})}
\left(\frac{1+z_\mathrm{q}}{1+z_j}\right)^{-\alpha+1}
\left(\frac{d_\mathrm{L}(z_\mathrm{q},0)}{d_\mathrm{L}(z_\mathrm{q},z_j)}\right)^2\nonumber\\
& &\times\int_1^\infty x^{-\alpha-4}\mathrm{exp}\left(-\sum_{k=1}^{j-1}N_{i,k}\sigma_i x^{-3}
\left(\frac{1+z_k}{1+z_j}\right)^{-3}\right)\,\mathrm{d}x,
\end{eqnarray}
with the photoionisation cross section at the Lyman limit $\sigma_i$, the
observed Lyman limit flux $f_{\nu,i}$ and $x=\nu/\nu_i$ with the Lyman limit
frequency $\nu_i$. Extrapolating the power law continuum flux to the
\ion{He}{ii} Lyman limit yields
$f_{\nu,\ion{He}{ii}}=f_{\nu,\ion{H}{i}}4^{-\alpha}$. With the spectral index
$\alpha$ from Table~\ref{lylimitfluxes} we obtain $\eta_\mathrm{min}\simeq 2.3$
for QSO~C.

We simulated $\eta(z)$ for a set of 1000 Monte Carlo Ly$\alpha$ forest spectra
generated with the procedure discussed in Sect.~\ref{forestratio_los}. We assumed
$\Gamma_{\ion{H}{i},\mathrm{b}}=1.75\times 10^{-12}$~s$^{-1}$ corresponding to
the UV background from Sect.~\ref{trprox_h1} and $\eta_\mathrm{b}=100$, which
agrees with the median $\eta$ towards HS~1700$+$6416 in the redshift range under
consideration \citep{fechner06}. The intervening absorbers successively block
the quasar flux. Especially, every absorber with $\log{N_\ion{H}{i}}>15.8$ will
truncate the quasar flux at $h\nu>4$~ryd due to a \ion{He}{ii} Lyman limit
system, leading to an abrupt softening of the radiation field.

\begin{figure}
\resizebox{\hsize}{!}{\includegraphics[angle=270]{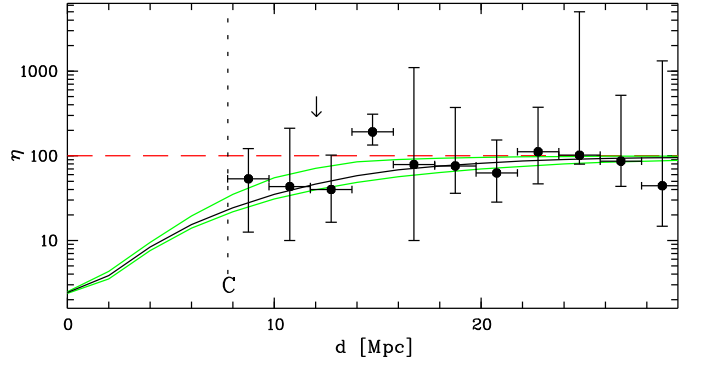}}
\caption{\label{he2347_etamodel} Column density ratio $\eta$ vs. proper distance
$d$. The black line shows the modelled decrease of the median $\eta$ approaching
QSO~C with respect to the ambient $\eta_\mathrm{b}=100$ (dashed line). Green
lines mark the upper and lower quartiles of the simulated $\eta$ distribution in
bins of $\Delta d=2$~Mpc. QSO~C is located at $7.75$~Mpc. Filled circles show the
median $\eta$ from FR07 in concentric rings of $\Delta d=2$~Mpc around the
quasar. Error bars are the quartile distances to the median. The arrow marks the
metal line system at $z=2.7122$ at $d=12.03$~Mpc. [See the online edition of the
Journal for a colour version of this figure.]}
\end{figure}

Figure~\ref{he2347_etamodel} presents the simulated decrease of the median
$\eta$ approaching QSO~C assuming a constant quasar luminosity, isotropic
radiation and an infinite quasar lifetime together with the upper and lower
percentiles of the $\eta$ distribution obtained in bins of proper distance
$\Delta d=2$~Mpc. The spread in the simulated $\eta$ is due to line-of-sight
differences in the absorber properties. Since we consider the transverse
proximity effect, we are limited to a proper distance $d\ga 7.75$~Mpc
(Table~\ref{lylimitfluxes}). The model agrees reasonably with the median $\eta$
of the data obtained in concentric rings around the quasar. As expected,
individual $\eta$ values strongly deviate from this simple model due to the
assumptions of the quasar properties (constant luminosity and spectral index,
isotropic radiation) and due to the unknown real distribution of absorbers in
transverse direction. Recently, \citet{hennawi07} found evidence for excess
small-scale clustering of high-column density systems in transverse direction
to quasar sightlines. In Sect.~\ref{hardness_qsoc} we found indications that
the $\eta$ distribution around QSO~C is not symmetric, which could be due to
such anisotropic shielding. However, this does not imply an intrinsic
anisotropy due to the unknown matter distribution around the quasar and the
large uncertainties in individual $\eta$ values. Moreover, the line-of-sight
variance at a constant $\eta=100$ is too small to explain the large observed
spread of the $\eta$ values. Clearly, a self-consistent explanation of the
small-scale $\eta$ fluctuations would require hydrodynamical simulations of
cosmological radiative transfer in order to investigate possible shielding
effects and the statistical distribution of $\eta$ values near quasars. While
there is recent progress in case of the UV background
\citep{sokasian03,croft04,maselli05,bolton06}, a proper treatment of
three-dimensional radiative transfer in the IGM around a quasar is still in its
infancy. However, our simplified approach suggests that QSO~C is capable of
changing the spectral shape of the UV radiation field by the right order of
magnitude to explain the low $\eta$ values in its vicinity. Also a variation in
the sizes and the centres of the bins chosen for Fig.~\ref{he2347_etamodel} does
not drastically change the indicated excess of low $\eta$ at $d\la 14$~Mpc.
Figure~\ref{he2347_etamodel} also shows very clearly that the sphere of
influence for the transverse proximity effect is not limited to the immediate
vicinity of the quasar.

\begin{figure}
\resizebox{\hsize}{!}{\includegraphics{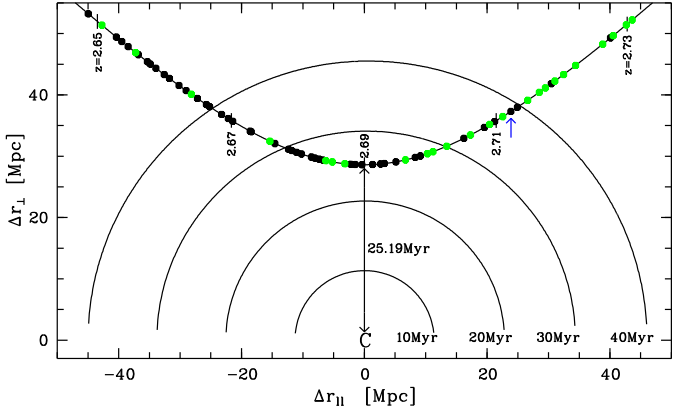}}
\caption{\label{he2347_isotime} Transverse comoving separation $\Delta r_\perp$
vs. line-of-sight comoving separation $\Delta r_\parallel$ with respect to
QSO~C. Black (green) points denote absorbers with $\eta<100$ ($\eta\ge 100$) on
the line of sight towards HE~2347$-$4342 (curved line) with indicated redshifts.
The blue arrow points to the metal line system at $z=2.712$. The half circles
show the distance travelled by light emitted at the indicated times prior to our
observation. The minimum light travel time between the two lines of sight is
$25.19$~Myr. [See the online edition of the Journal for a colour version of this
figure.]}
\end{figure}

Figure~\ref{he2347_isotime} shows a two-dimensional cut in comoving space near
QSO~C in the plane spanned by both lines of sight. The minimum separation of
both lines of sight corresponds to a light travel time of $\simeq 25.2$~Myr,
but the lifetime of QSO~C could be $\ga 40$~Myr due to the low $\eta$ values at
larger distances. The fluctuations of the UV spectral shape could be explained
by shadowing of the hard QSO radiation by unknown intervening structures between
both lines of sight.

\subsection{Other regions with an inferred hard UV radiation field}
\label{hardrad_noqso}

\begin{figure}
\resizebox{\hsize}{!}{\includegraphics{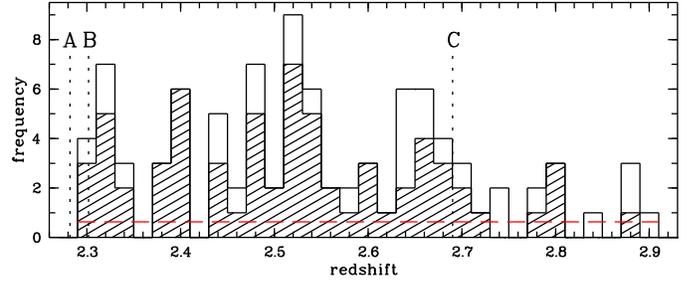}}
\caption{\label{he2347_etahist} Redshift distribution of low-$\eta$ absorbers.
The open (hashed) histogram shows all ($N_\ion{H}{i}\le 10^{14}$~cm$^{-2}$)
absorbers with $\eta\le 10$. Letters and dotted lines mark foreground quasars.
The horizontal dashed line denotes the estimated average number of absorbers
scattered from $\eta=80$ to $\eta\le 10$ ($\simeq 0.64$ per bin).}
\end{figure}

In Fig.~\ref{he2347forestratio} we note two additional regions at $z\sim 2.48$
and $z\sim 2.53$ where $R$ is prominently small and where there is no nearby
quasar. Also the fitted $\eta(z)$ shows very low values apparently unrelated to
a known foreground quasar. Figure~\ref{he2347_etahist} displays the redshift
distribution of the $\eta\le 10$ subsample. The low $\eta$ values are clustered
with two peaks near the foreground quasars, but also at $z\sim 2.40$,
$z\sim 2.48$ and $z\sim 2.53$. At the first glance the existence of such regions
seems to undermine the relation between the foreground quasars and a low $\eta$
in their vicinity. However, there are several plausible explanations for the
remaining low $\eta$ values:
\begin{enumerate}
\item{\emph{Unknown quasars:} We can
conclude from Paper~I that the quasars responsible for hardness fluctuations may
be very faint (like \object{Q~0302-D113} in Paper~I) or may reside at large
distances (\object{Q~0301$-$005} in Paper~I). QSO~C is located near the edge of
our survey area centred on HE~2347$-$4342, so other quasars capable of
influencing the UV spectral shape might be located outside the field of view.
Moreover, in order to sample the full quasar luminosity domain ($M_B\le -23$) at
$z\sim 2.5$ our survey is still too shallow by $\sim 1$~magnitude. Therefore,
a larger and/or deeper survey around HE~2347$-$4342 is desirable.}
\item{\emph{Quasar lifetime:}
Assuming that quasars are long-lived and radiate isotropically, every
statistically significant low $\eta$ fluctuation should be due to a nearby
quasar. On the other hand, short-lived quasars will not be correlated with a
hard radiation field due to the light travel time from the quasar to the
background line of sight. Quasar lifetimes are poorly constrained by
observations to $1\la t_\mathrm{q}\la 100$~Myr \citep{martini04}. This could be
short enough to create relic light echoes from extinct quasars. The comoving
space density of quasars with $M_B<-23$ at $z\simeq 2.5$ is
$\simeq 3.7\times 10^{-6}$~Mpc$^{-3}$ \citep{wolf03} resulting in an average
proper separation of $\sim 18.5$~Mpc between two lines of sight. This translates
into a light travel time of $\sim 60$~Myr which is of the same order as the
quasar lifetime. So it is quite possible that some quasars have already turned
off, but their hard radiation is still present.}
\item{\emph{Obscured quasars:} Anisotropic emission of type~I quasars may lead
to redshift offsets between regions with an inferred hard radiation field and
quasars close to the line of sight. In the extreme case the putative quasar
radiates in transverse direction, but is obscured on our line of sight (type~II
quasar). The space density of type~II AGN at $z>2$ is very uncertain due to the
challenging optical follow-up that limits the survey completeness
\citep[e.g.][]{barger03,szokoly04,krumpe07}. Thus, the fraction of obscured AGN
at high redshift is highly debated \citep{akylas06,treister06}, but may well
equal that of type~I AGN in the luminosity range of interest \citep{ueda03}.}
\end{enumerate}

We believe that a combination of the above effects is responsible for the loose
correlation between low $\eta$ values and active quasars. In particular, at
$z\sim 2.4$ we infer a hard radiation field in a \ion{H}{i} void (FR07), which
may have been created by a luminous quasar that is unlikely to be missed by our
survey ($V\la 22$).

In Fig.~\ref{he2347_etahist} we also indicate the error level due to inaccurate
line fitting and noise in the \ion{He}{ii} data (dashed line) obtained from
simulated data (see below). The low number of $\eta$ values scattered from a
simulated $\eta=80$ to $\eta\le 10$ implies that the overdensities of such small
$\eta$ values are statistically significant. Constraining the sample to lines
with $\log(N_\ion{H}{i})<14$ due to a possible bias caused by thermal broadening
does not remove the significant clusters of lines with small $\eta$.

\subsection{Uncertainties in the spectral hardness}
\label{uncertainties}

Our findings are likely to be affected by random errors and possibly also by
systematic errors mostly related to the \ion{He}{ii} data. The poor quality of
the FUSE spectrum of HE~2347$-$4342 ($S/N\la 5$) contributes to the fluctuations
in $\eta$ even if the $\eta$ value was constant \citep[][see also below]{fechner06}.
The optical depth ratio $R$ should be less affected by noise, since it is an
average over a broader redshift range $\Delta z=0.005$. The low $S/N$ and the
generally high absorption at $\eta\gg 1$ provide uncertainty for the continuum
determination in the \ion{He}{ii} spectrum. The extrapolated reddened power law
is certainly an approximation.

Although the $\eta$ fitting results from FR07 are broadly consistent with
those of Z04 and agree well in the regions near the foreground quasars, there
are substantial differences in some redshift ranges. This is probably due to the
combined effects of low \ion{He}{ii} data quality, different data analysis
software and ambiguities in the deblending of lines. At present, $\eta$ cannot
be reliably determined at individual absorbers unless metal transitions provide
further constraints.

In order to assess the random scatter in $\eta$ due to the low $S/N$ \ion{He}{ii}
data and ambiguities in the line deblending of both species, we again used Monte
Carlo simulations. Ten \ion{H}{i} Ly$\alpha$ forest spectra were generated in
the range $2<z<3$ via the Monte Carlo procedure outlined in
Sect.~\ref{forestratio_los}. The resolution $R\sim 42000$ and $S/N=100$ closely
resembles the optical data of HE~2347$-$4342. We also generated the corresponding
\ion{He}{ii} forests at FUSE resolution and $S/N=4$. We assumed pure non-thermal
line broadening and $\eta=80$. Voigt profiles were automatically fitted to the
\ion{H}{i} spectra using AUTOVP
\footnote{http://ursa.as.arizona.edu/\textasciitilde rad/autovp.tar} \citep{dave97}.
The \ion{He}{ii} spectra were then automatically fitted with redshifts $z$ and
non-thermal Doppler parameters $b_\ion{H}{i}$ fixed from the fitted \ion{H}{i}
line lists, yielding 7565 simulated $\eta$ values. On average the recovered
$\eta$ is slightly higher than the simulated one (median $\eta\simeq89$) with a
large spread ($0<\eta\la 8000$), but only 285 lines have $\eta\le 10$. Thus, we
estimate a probability $P\simeq 3.8$\% that $\eta$ is scattered randomly from
$\eta=80$ to $\eta\le 10$ if the assumption of non-thermal broadening is
correct. Note that this probability is likely an upper limit due to the fact
that only \ion{H}{i} Ly$\alpha$ was used to obtain the line parameters, which
results in large error bars for saturated lines on the flat part of the curve of
growth. In the real data, these errors were avoided by fitting unsaturated
higher orders of the Lyman series wherever possible.

In the line sample by FR07, 94 out of the 526 absorbers have $\eta\le 10$,
whereas our simulation implies that only $\sim 20$ are expected to be randomly
scattered to $\eta\le 10$ if $\eta$ was constant. Thus, the major part of the
scatter of $\eta$ in the data is due to real fluctuations in the UV spectral
shape. The majority of the low $\eta\le 10$ values is inconsistent with
$\eta\ge 80$, so they indicate a hard radiation field in spite of the low $S/N$
in the \ion{He}{ii} data. Yet, due to the large intrinsic scatter obtained from
the simulations, individual $\eta$ values hardly trace the variations of the UV
spectral shape. Local spatial averages should be more reliable (FR07). Since
the transverse proximity effect zones always extend over some redshift range,
this requirement is fulfilled and \emph{on average} we reveal a harder radiation
field than expected.

Concerning the high tail of the simulated distribution at $\eta=80$,
$\sim 15$\% of the lines are returned with $\eta\ga 200$. This may indicate that
a fraction of the observed high $\eta$ values is still consistent with a
substantially harder radiation field, underlining that single $\eta$ values
poorly constrain the spectral shape.

Possibly, some $\eta$ values are systematically too low due to the assumption of
non-thermal broadening ($b_\ion{He}{ii}=b_\ion{H}{i}$) when fitting the
\ion{He}{ii} forest. FR07 found that this leads to underestimated $\eta$ values
at $N_\ion{H}{i}\ga 10^{13}$~cm$^{-2}$ if the lines are in fact thermally
broadened ($b_\ion{He}{ii}=0.5b_\ion{H}{i}$). Non-thermal broadening is caused
by turbulent gas motions or the differential Hubble flow, with the latter
affecting in particular the low-column density forest. Thermal broadening
becomes important in collapsed structures at high column densities. In
simulations of the Ly$\alpha$ forest, non-thermal broadening has been found to
dominate \citep{zhang95,zhang98,hernquist96,weinberg97,bolton06,liu06}. This has
been confirmed observationally for the low-column density forest
\citep[Z04;][]{rauch05}. On the other hand, eight out of eleven absorbers with
$N_\ion{H}{i}> 10^{14}$~cm$^{-2}$ in the vicinity of QSO~C have $\eta\le 10$
(Fig.~\ref{he2347etaplot_c}). Although the column density ratio of these
absorbers could be underestimated due to an unknown contribution of thermal
broadening, the statistical evidence for a hard radiation field is based on the
vast majority of low-column density lines. The median $\eta$ obtained in this
region does not increase significantly after excluding the suspected lines
($\sim 53$ vs.\ $\sim 40$). This is still much lower than the median
$\eta\sim 100$ towards HS~1700$+$6416 in this redshift range \citep{fechner06}.
Therefore, it is unlikely that our results are biased due to the assumed line
broadening.

\section{Conclusions}\label{conclusions}

Traditionally, the transverse proximity effect of a quasar has been claimed to
be detectable as a radiation-induced void in the \ion{H}{i} Ly$\alpha$ forest.
But due to several systematic effects like quasar variability, finite quasar
lifetime, intrinsic overdensities around quasars, or anisotropic radiation, most
searches yielded negative results \citep[e.g.][]{schirber04,croft04}.

In this paper, we have analysed the fluctuating spectral shape of the UV
background in the projected vicinity of the three foreground quasars
QSO~J23503$-$4328, QSO~J23500$-$4319 and QSO~J23495$-$4338 (dubbed QSO~A, B and
C) on the line of sight towards HE~2347$-$4342 ($z=2.885$). By comparing the
\ion{H}{i} absorption and the corresponding \ion{He}{ii} absorption, we have
presented evidence for a statistical excess of hard UV radiation near the
foreground quasars. However, due to contamination of the forests near QSO~A
($z=2.282$) and QSO~B ($z=2.302$), the evidence is strongest for QSO~C
($z=2.690$). We interpret these
indicators for an excess of hard radiation near the foreground quasars as a
manifestation of the transverse proximity effect. A simple model indicates that
the foreground quasars are capable of generating the observed hard radiation
over the observed distances of several Mpc.
Furthermore, we tried to model the ionising radiation field of three metal line
systems close to the foreground quasars. Two of those are strongly affected by
line blending and do not allow for reliable photoionisation models. The
remaining system can be modelled reasonably with or without a contribution by a
local quasar. Future larger samples of highly ionised unblended metal systems
near foreground quasars may provide evidence for local hardness fluctuations.

In \citet{worseck06} we revealed the transverse proximity effect as a systematic
local hardness fluctuation around four foreground quasars near Q~0302$-$003 and
pointed out that the relative UV spectral hardness is a sensitive physical
indicator of the proximity effect over distances of several Mpc. In this study
we are able to confirm this on a second line of sight. Evidently, small-scale
transmission fluctuations in the \ion{H}{i} forest can dilute the small
predicted signature of the effect. However, the hard spectral shape of the UV
radiation field still indicates the transverse proximity effect despite the
\ion{H}{i} density fluctuations. Thus, we confirm our previous result that the
spectral hardness breaks the density degeneracy that affects the traditional
searches for the proximity effect. Moreover, the predicted transverse proximity
effect of the quasars in the \ion{H}{i} forest is weak due to the high UV
background at 1~ryd. Still the UV spectral shape is able to discriminate local
UV sources independent of the amplitude of the UV background.

\citet{bolton06} find that the large UV spectral shape fluctuations in the IGM
are likely due to the small number of quasars contributing to the \ion{He}{ii}
ionisation rate at any given point, whereas the \ion{H}{i} ionisation rate is
rather homogeneous due to the probable contribution of star-forming galaxies
\citep[e.g.][]{bianchi01,sokasian03,shapley06}. Our findings confirm the picture
that AGN create the hard part of the intergalactic UV radiation field. If the
quasar is active long enough, its hard radiation field can be observed
penetrating a background line of sight. It is also likely that light echoes from
already extinguished quasars are responsible for some locations of hard
radiation without an associated quasar. The transverse proximity effect of QSO~C
implies a minimum quasar lifetime of $\sim 25$~Myr (probably even $\sim 40$~Myr),
providing additional constraints to more indirect estimates
\citep[e.g.][and references therein]{martini04}.

However, the UV radiation field near the foreground quasars is not homogeneously
hard as naively expected, but still shows fluctuations. Apart from substantial
measurement uncertainties, the unknown density structure around the quasar could
shield the ionising radiation in some directions, maybe even preferentially in
transverse direction to the line of sight \citep{hennawi07}. Thus, radiative
transfer effects may become important to explain a fluctuating UV spectral shape
in the presence of a nearby quasar. Large-scale simulations of cosmological
radiative transfer with discrete ionising sources are required to adress these
issues in detail.

Moreover, the \ion{He}{ii} forest has been resolved so far only towards two
quasars at a very low $S/N\la 5$. While the low data quality primarily
creates uncertainties in the spectral shape on small spatial scales, large
scales could be affected by cosmic variance. Thus, the general redshift evolution
of the UV spectral shape is not well known and estimates obtained from single
lines of sight may well be biased by local sources.

\begin{acknowledgements}

We thank the staff of the ESO observatories La Silla and Paranal for their
professional assistance in obtaining the optical data discussed in this
paper. We are grateful to Peter Jakobsen for agreeing to publish the
quasars from his survey. We thank Gerard Kriss
and Wei Zheng for providing the reduced FUSE spectrum of HE~2347$-$4342. 
Tae-Sun Kim kindly supplied an additional line list of HE~2347$-$4342.
GW and ADA acknowledge support by a HWP grant from the state of Brandenburg,
Germany. CF is supported by the Deutsche Forschungsgemeinschaft under
RE 353/49-1. We thank the anonymous referee for helpful comments.

\end{acknowledgements}

\bibliography{7585}
\end{document}